\documentclass{sig-alternate-05-2015}

\usepackage{mathptmx}
\usepackage{graphicx}
\usepackage{times}
\usepackage{caption}
\usepackage{subfig}

\makeatletter
\newif\if@restonecol
\makeatother

\usepackage[algo2e,ruled,vlined]{algorithm2e} 
\usepackage{amssymb}
\usepackage{amsmath}

\usepackage{multirow}
\usepackage{tabularx}

\makeatletter
\def\@copyrightspace{\relax}
\makeatother

\begin{document}


\doi{XXXXXXXXXX}

\isbn{XXXXXXXXX}



%

\title{Automated capture and delivery of assistive task guidance with an eyewear computer: {T}he {G}laci{AR} system}

%
%
%
%
%

\numberofauthors{3} 
%
\author{
%
%
\alignauthor
Teesid Leelasawassuk\titlenote{csztl@bristol.ac.uk}\\
       \affaddr{University of Bristol}\\
\alignauthor
Dima Damen\titlenote{csxda@bristol.ac.uk}\\
       \affaddr{University of Bristol}\\
\alignauthor
Walterio Mayol-Cuevas\titlenote{wmayol@cs.bris.ac.uk}\\
       \affaddr{University of Bristol}\\
}


\maketitle
\begin{abstract}
In this paper we describe and evaluate a mixed reality system that aims to augment users in task guidance applications by combining automated and unsupervised information collection with minimally invasive video guides. The result is a self-contained system that we call GlaciAR (Glass-enabled Contextual Interactions for Augmented Reality), that  operates by extracting contextual interactions from observing users performing actions. GlaciAR is able to i) automatically determine moments of relevance based on a head motion attention model, ii) automatically produce video guidance information, iii) trigger these video guides based on an object detection method, iv) learn without supervision from observing multiple users and v) operate fully on-board a current eyewear computer (Google Glass).  We describe the components of GlaciAR together with evaluations on how users are able to use the system to achieve three tasks. We see this work as a first step toward the development of systems that aim to scale up the notoriously difficult authoring problem in guidance systems and where people's natural abilities are enhanced via minimally invasive visual guidance. 
\end{abstract}

%
%

\ccsdesc[500]{Human-centered computing~Mixed / augmented reality}
\ccsdesc[300]{Human-centered computing~HCI design and evaluation methods}

%
%

%
%
\printccsdesc


\keywords{Augmented Reality; Task Guidance; Eyewear computing}

\section{Introduction}

\begin{figure*}[ht!]
\centering
\includegraphics[width=0.27\columnwidth]{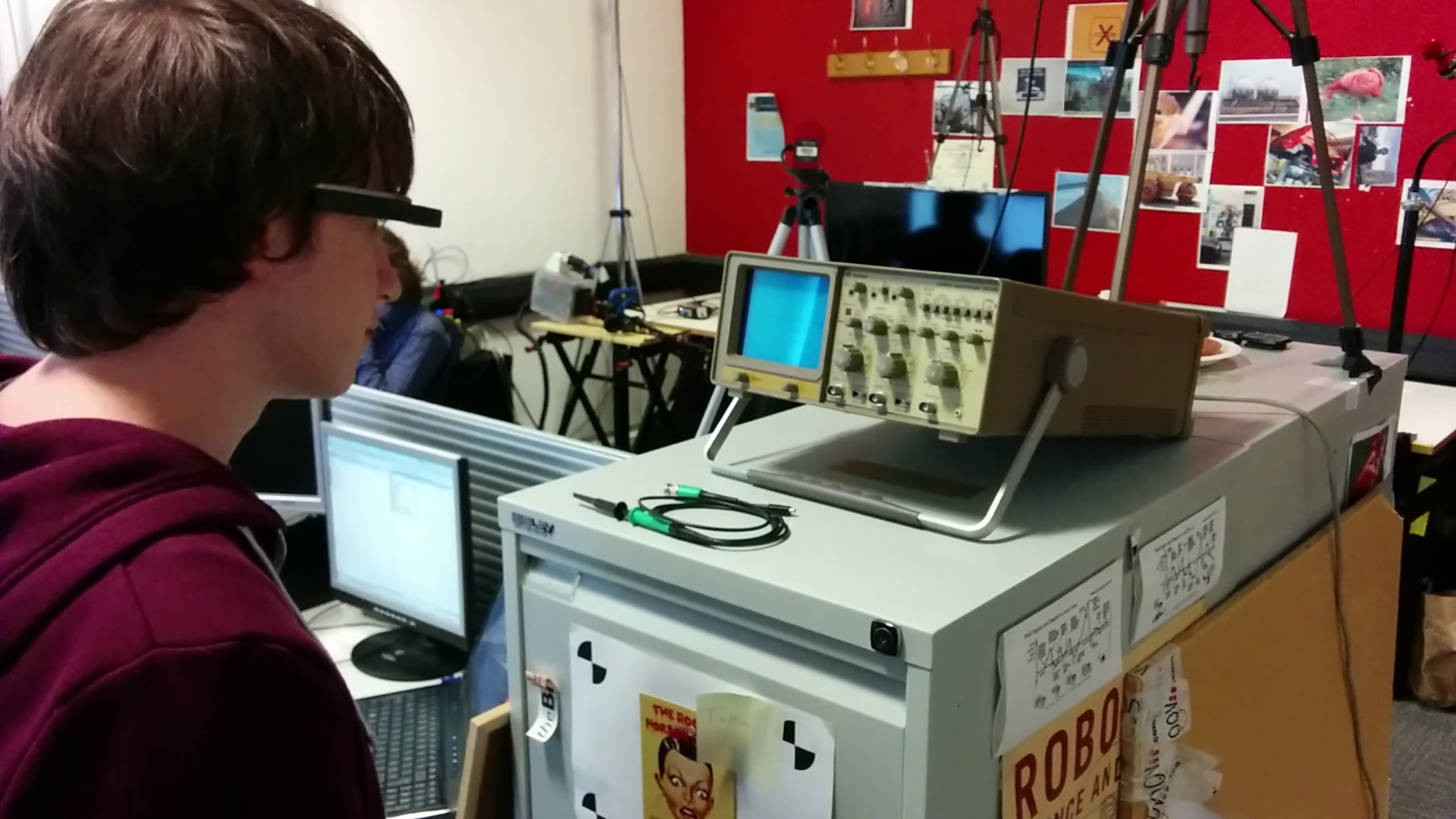}
\includegraphics[width=0.27\columnwidth]{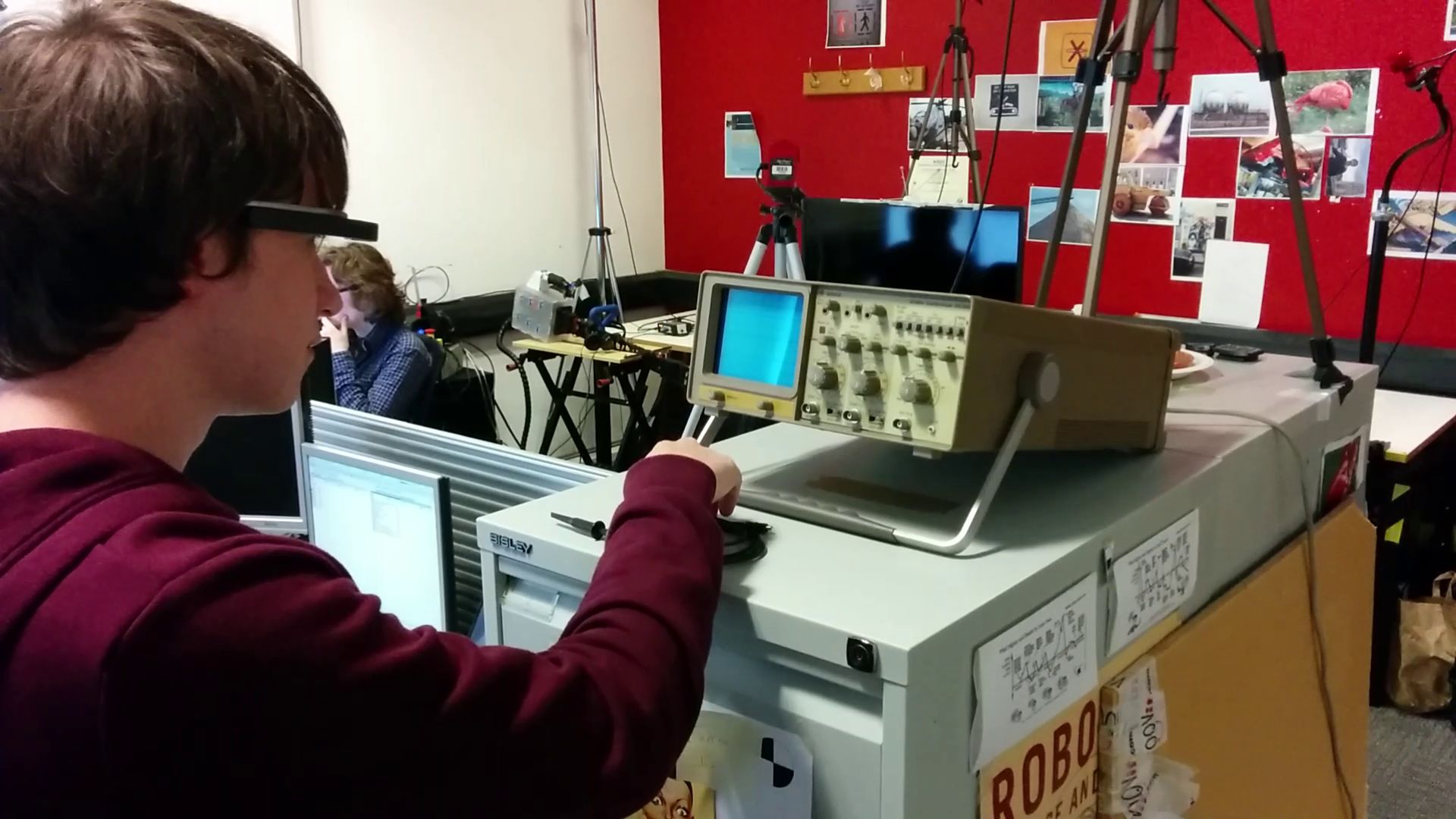}
\includegraphics[width=0.27\columnwidth]{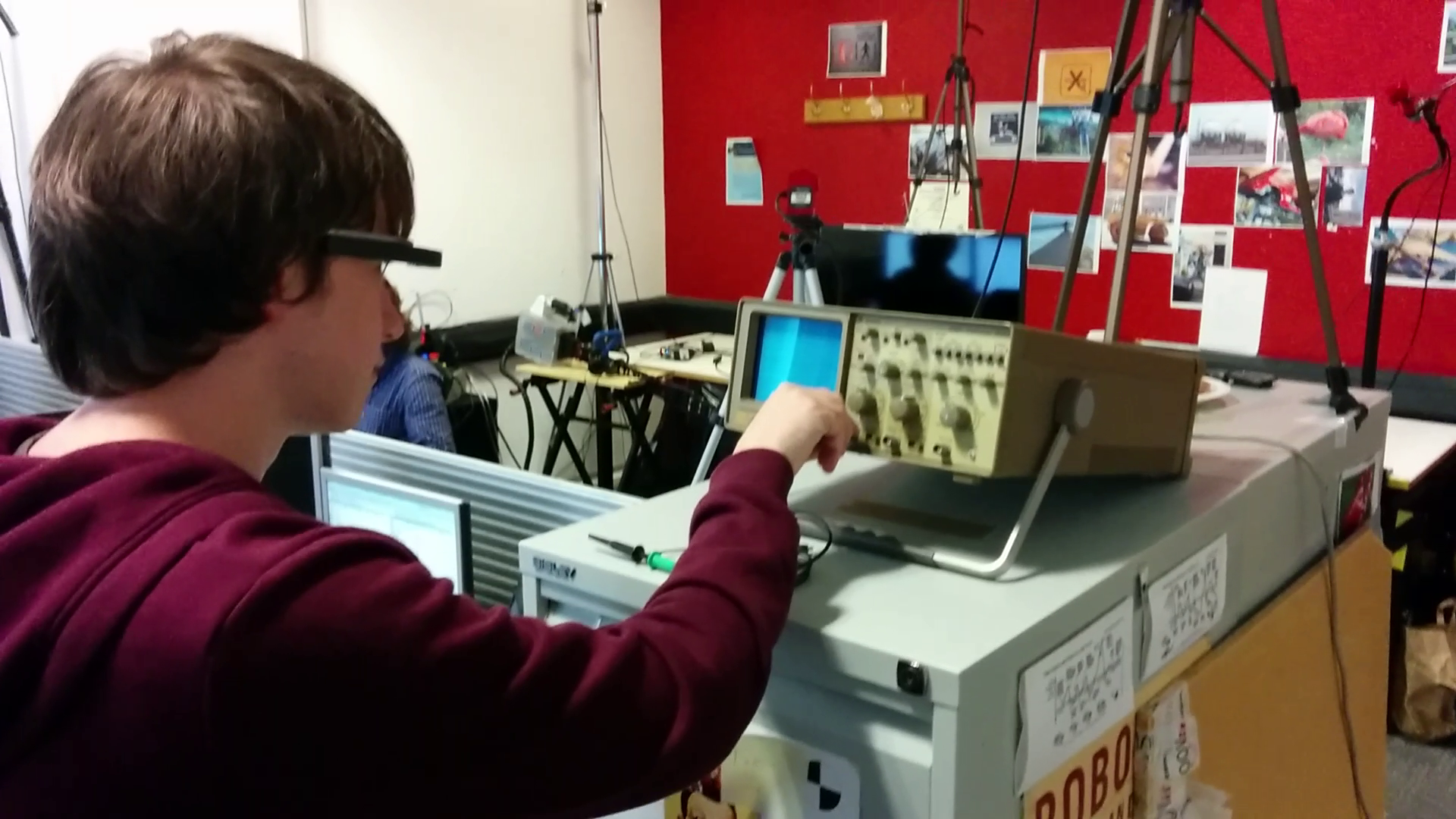}
\includegraphics[width=0.27\columnwidth]{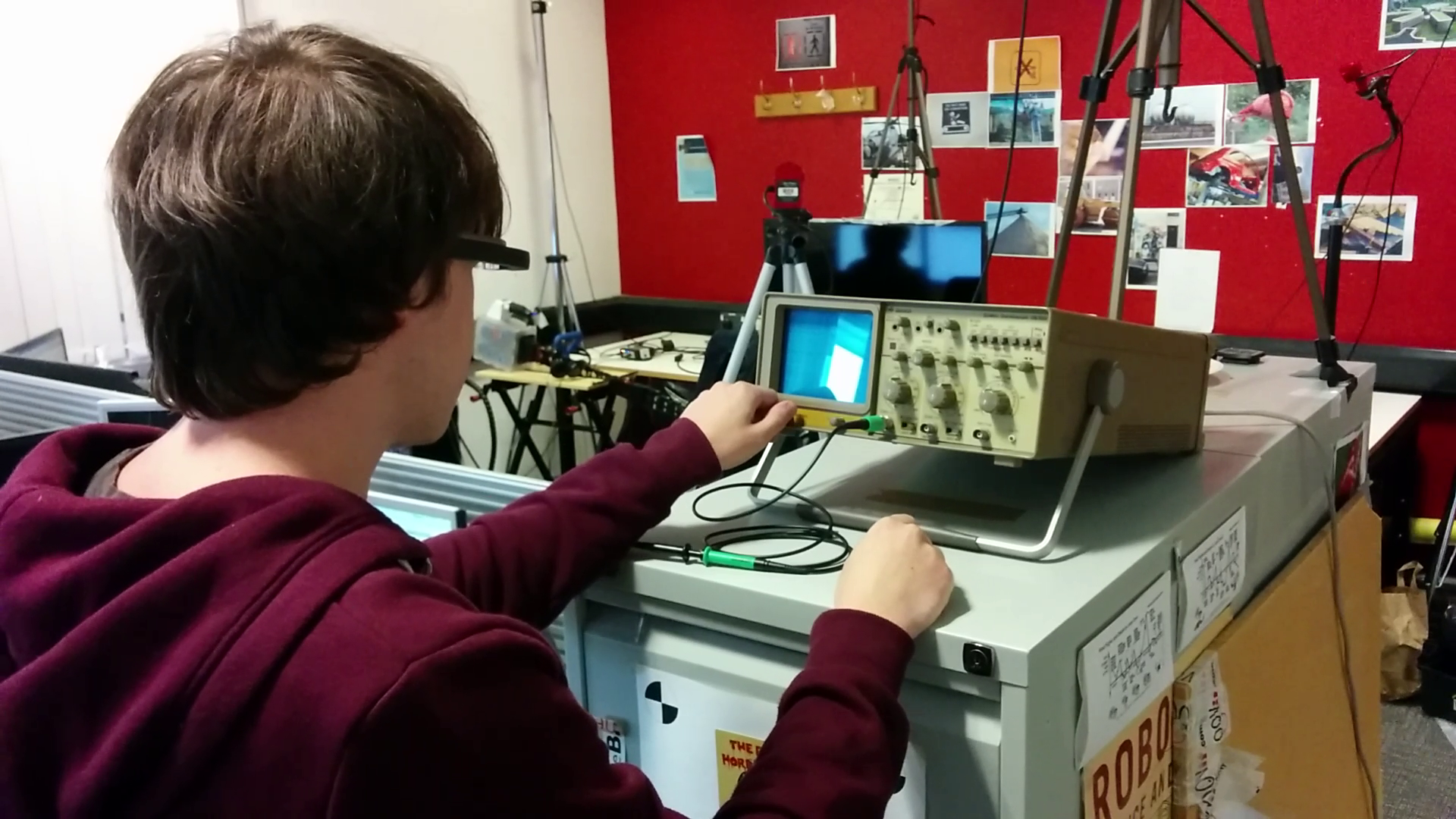}
\includegraphics[width=0.27\columnwidth]{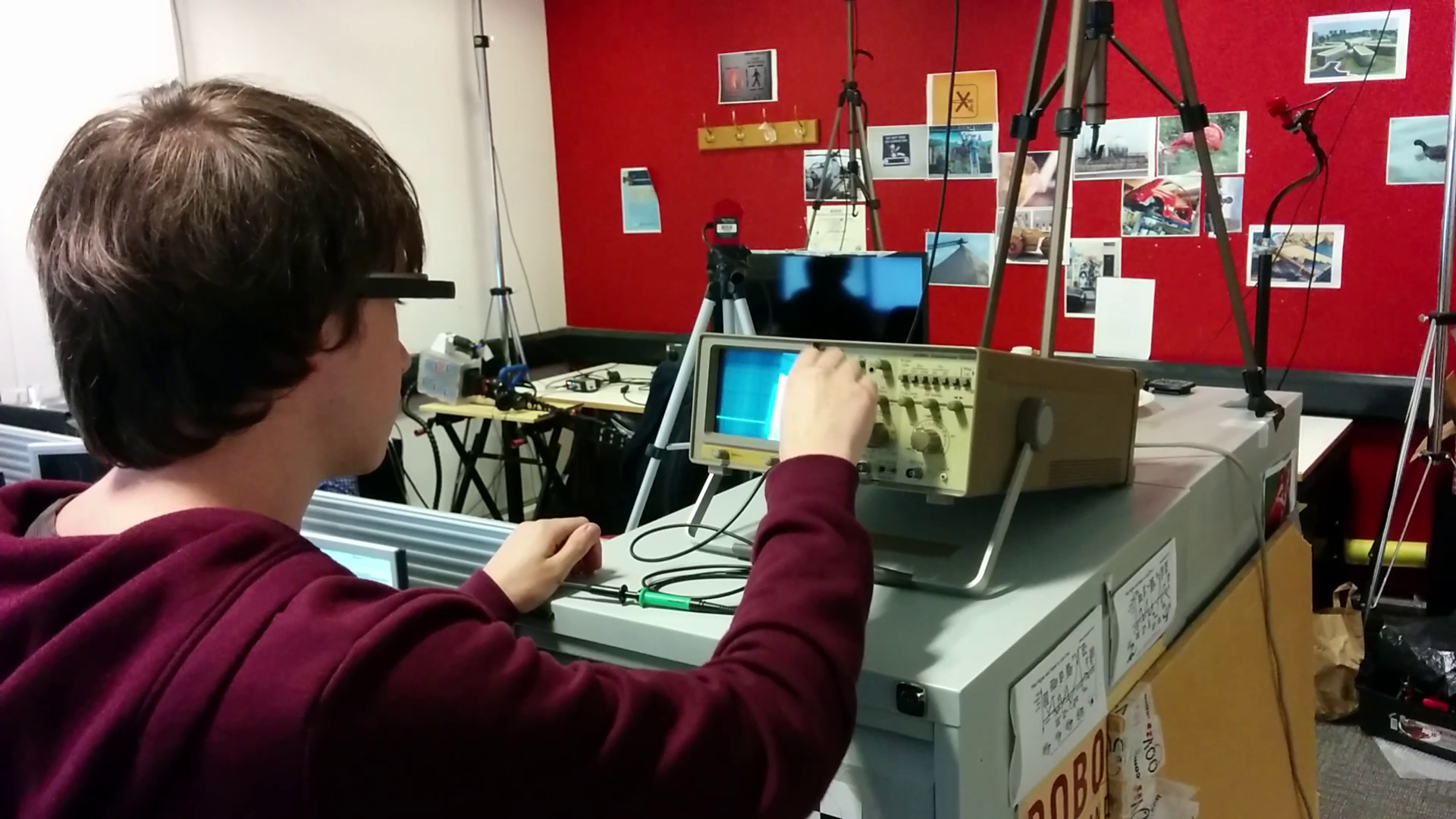}
\includegraphics[width=0.27\columnwidth]{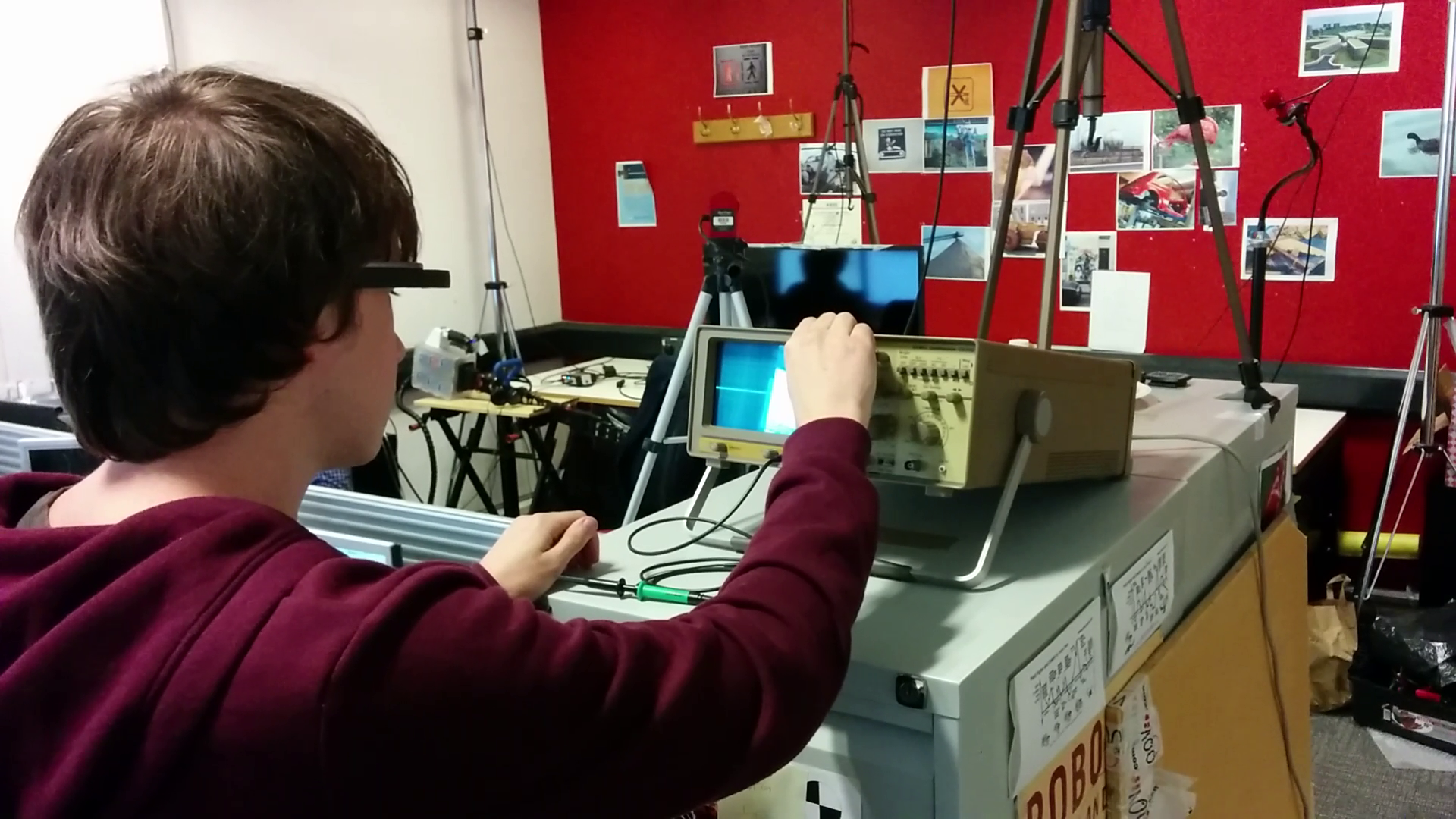}
\includegraphics[width=0.27\columnwidth]{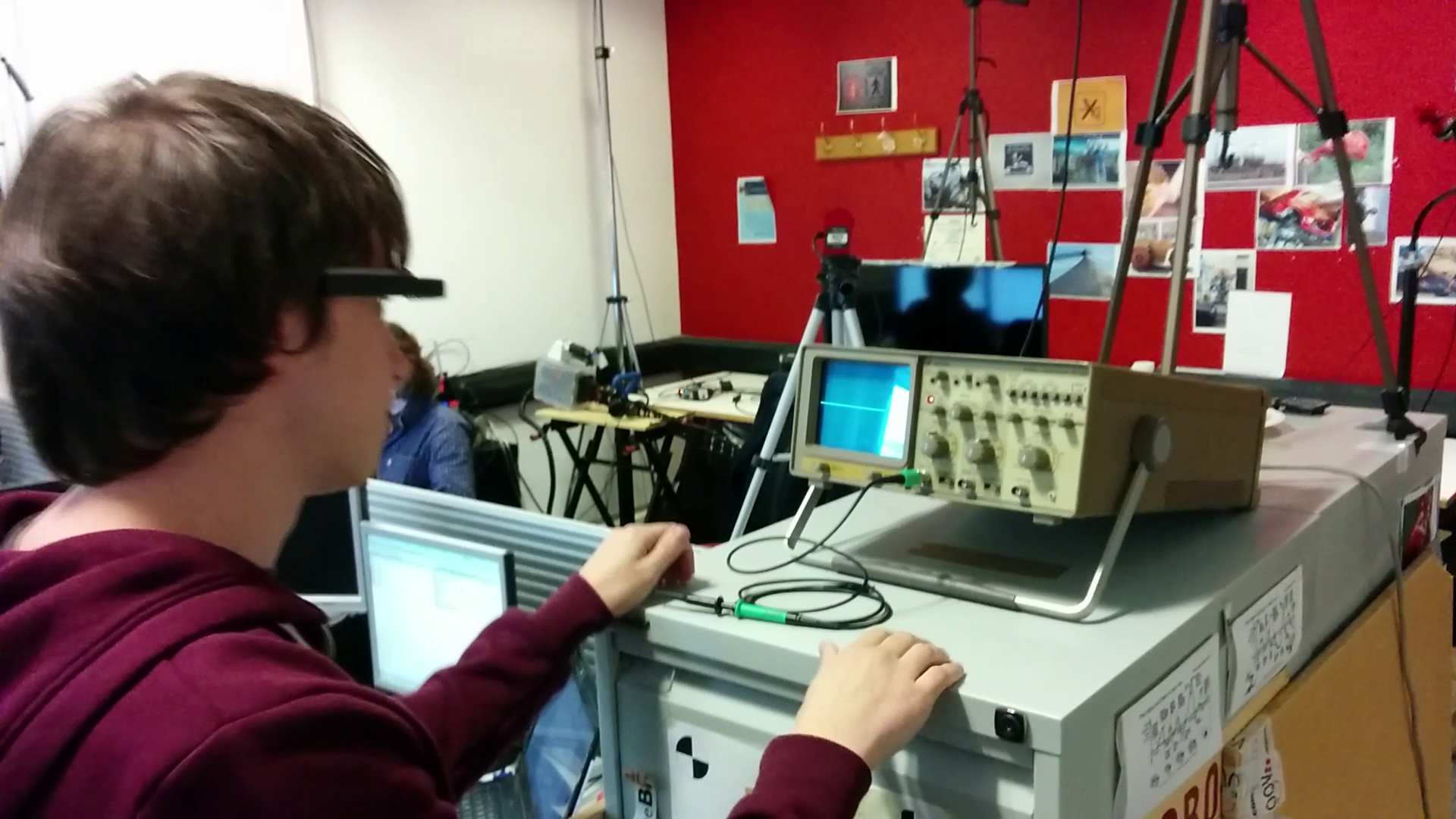}\\
\includegraphics[width=0.27\columnwidth]{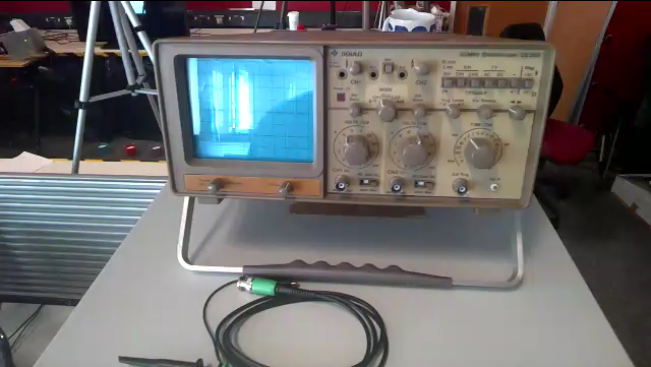}
\includegraphics[width=0.27\columnwidth]{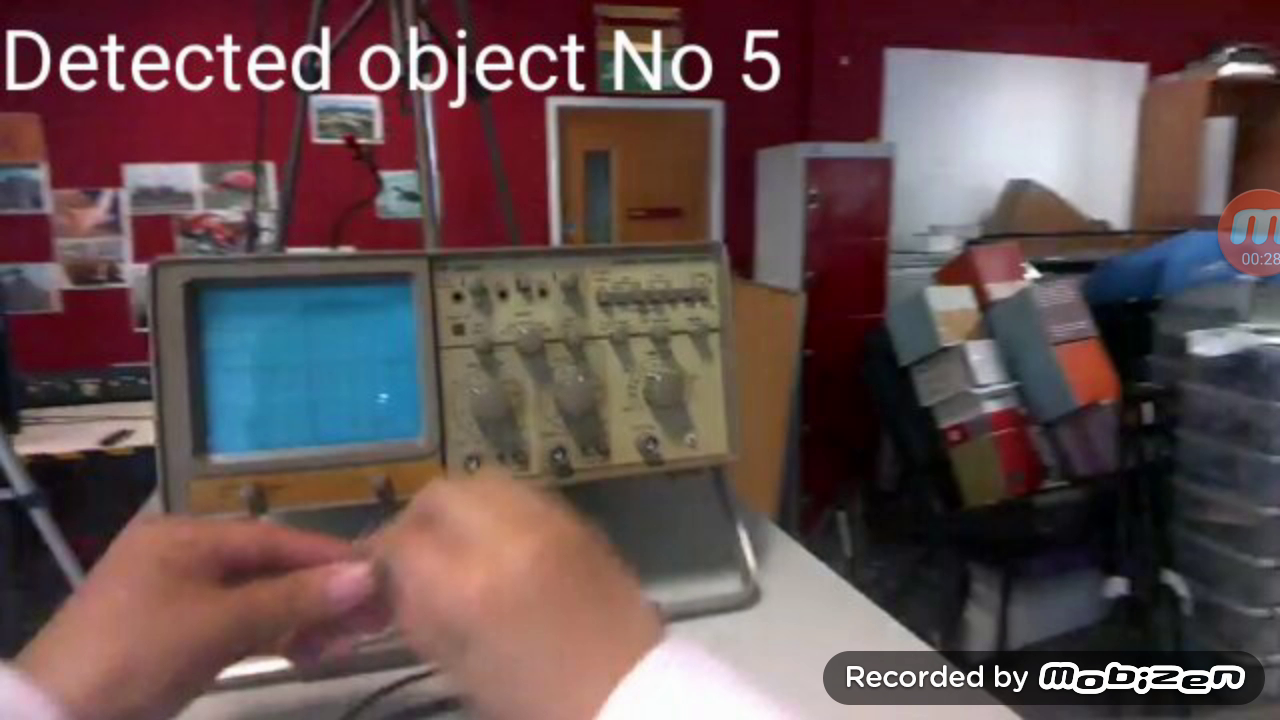}
\includegraphics[width=0.27\columnwidth]{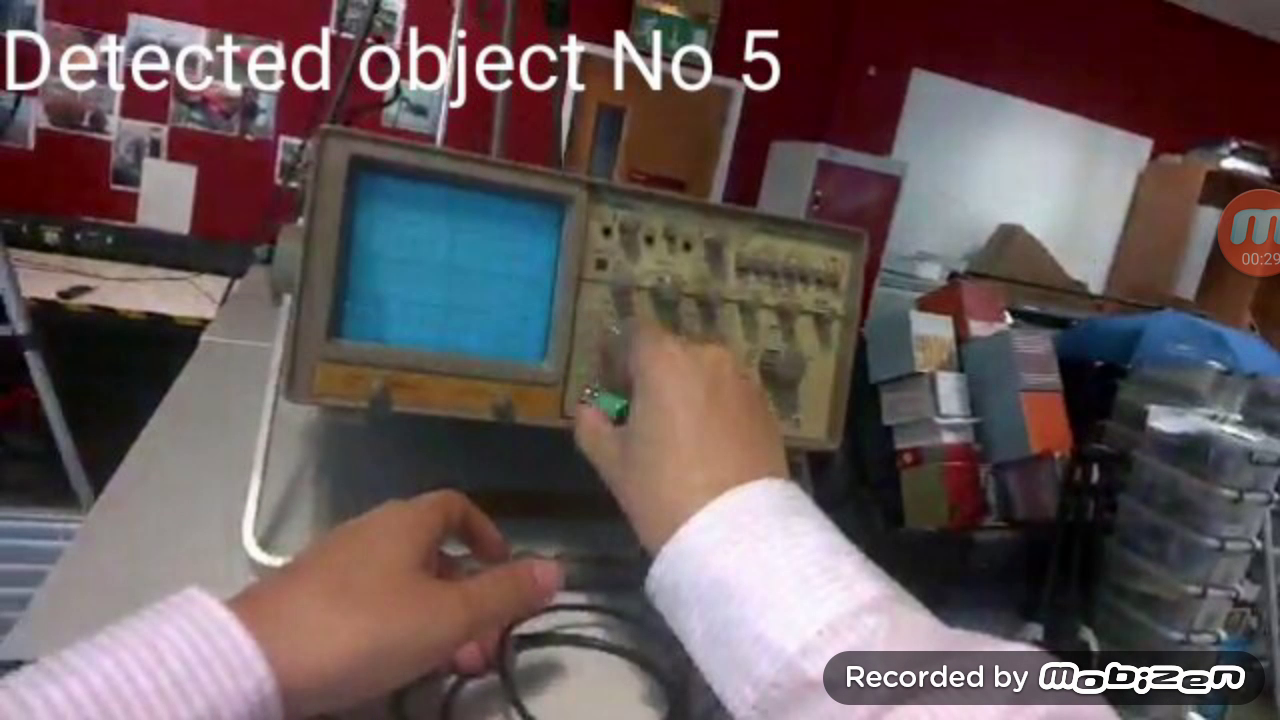}
\includegraphics[width=0.27\columnwidth]{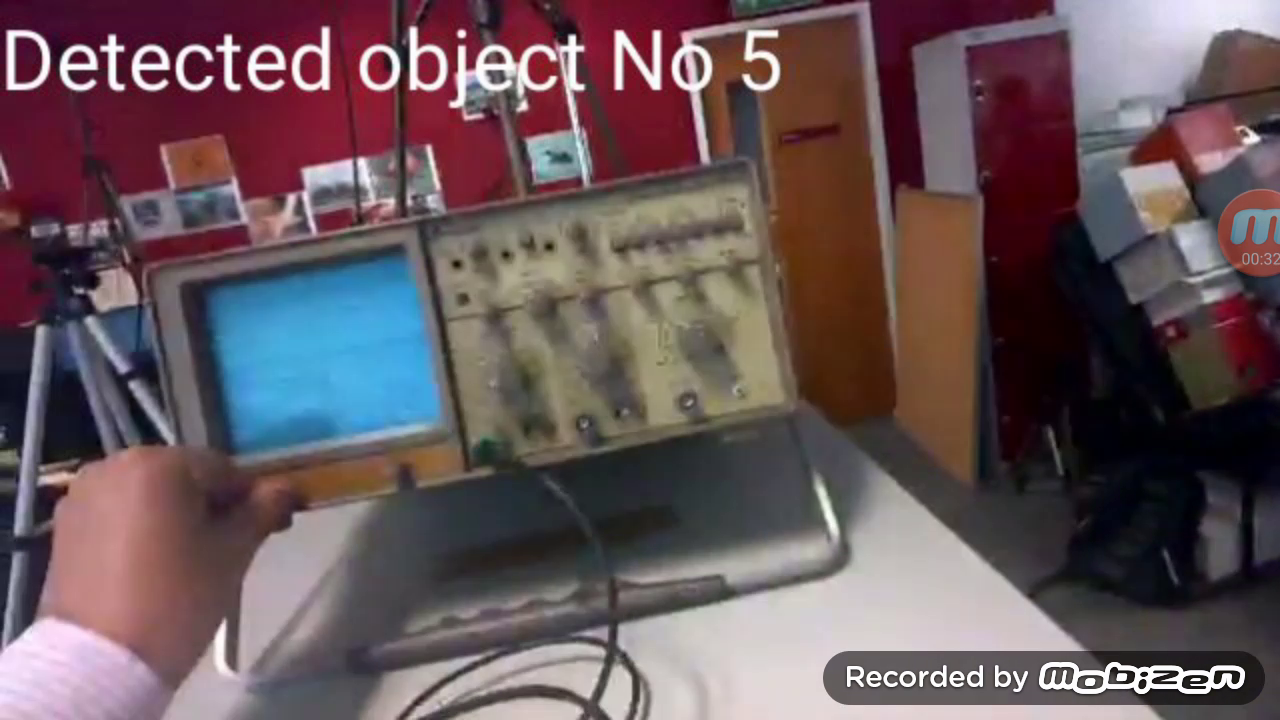}
\includegraphics[width=0.27\columnwidth]{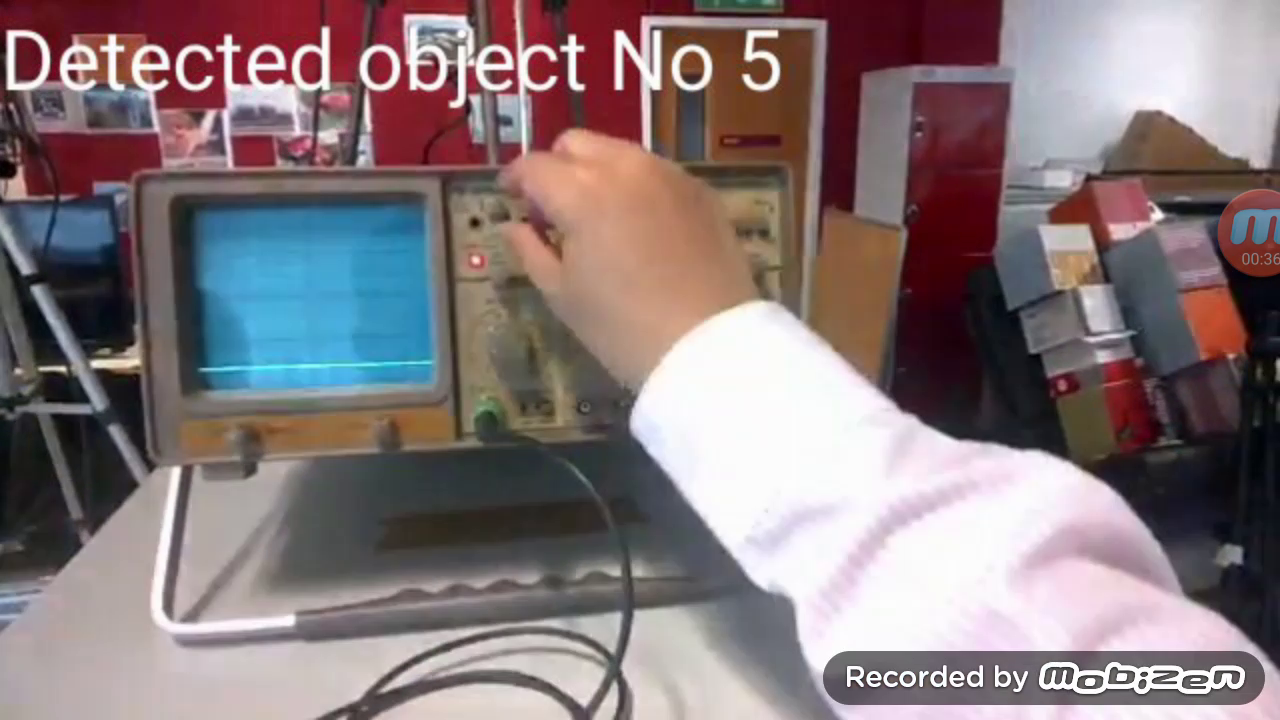}
\includegraphics[width=0.27\columnwidth]{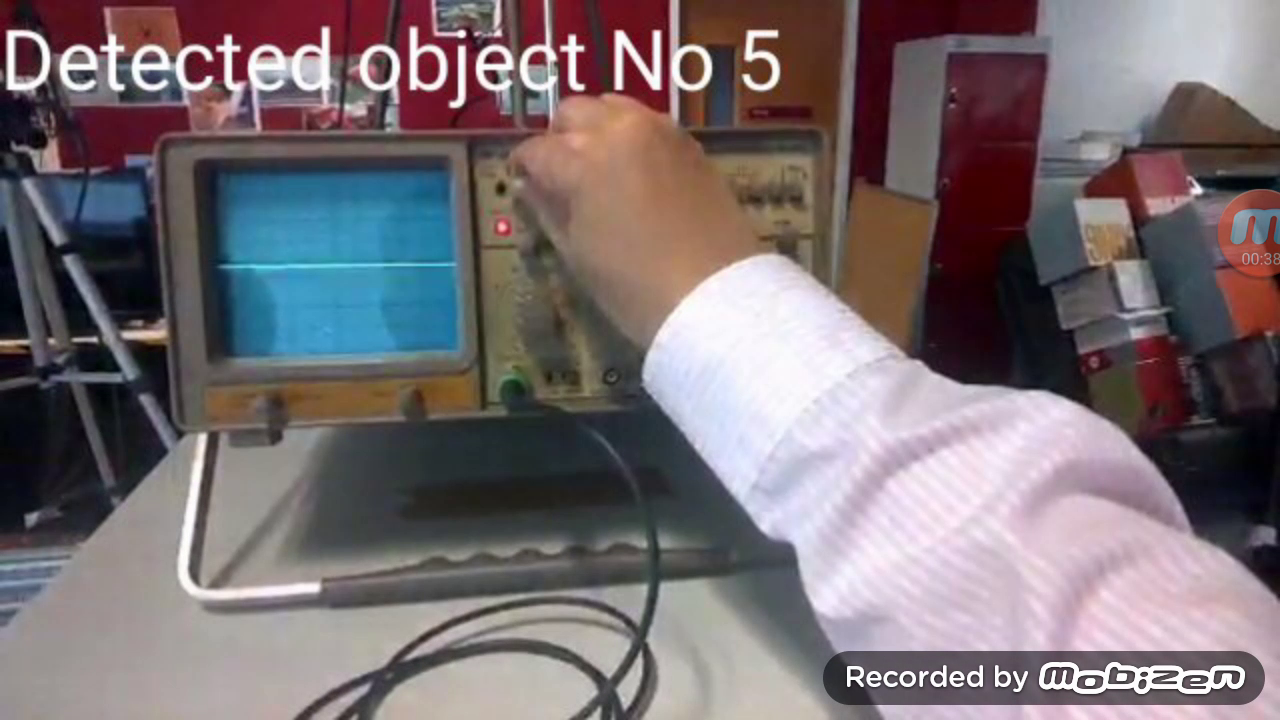}
\includegraphics[width=0.27\columnwidth]{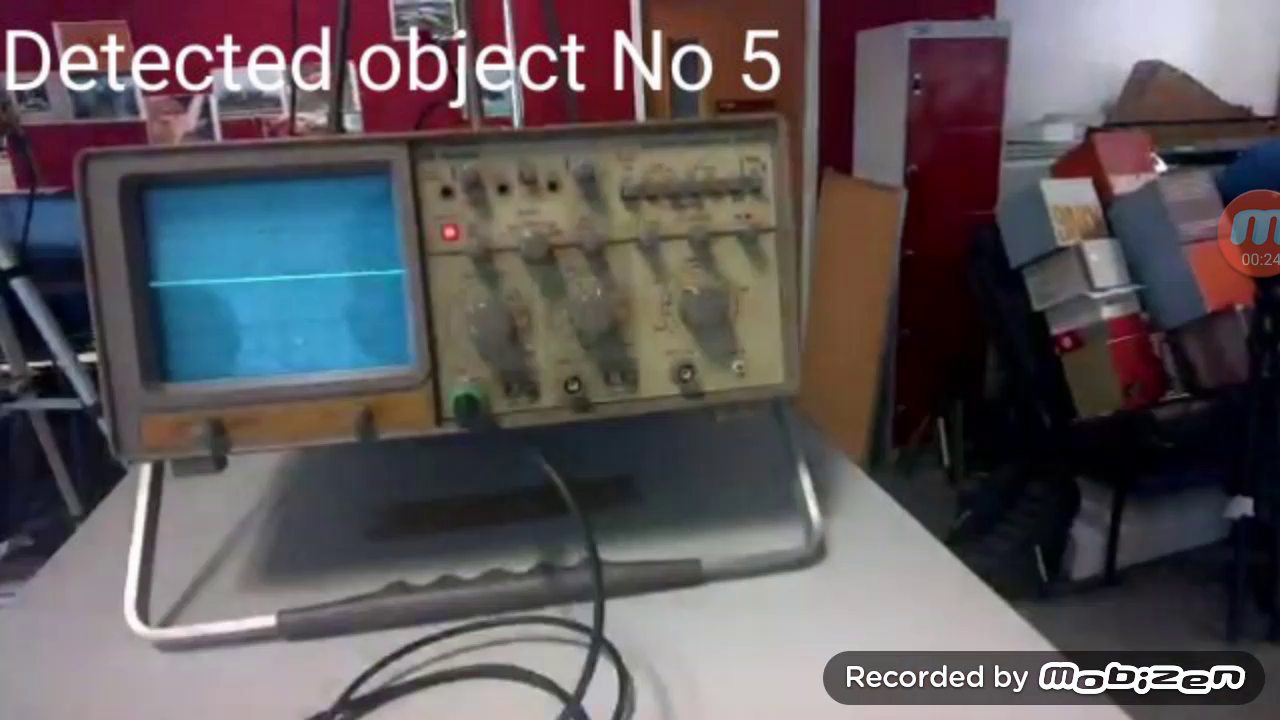}
\caption{A user of GlaciAR that has never before used an oscilloscope (top row) walks up to it and is intuitively guided to set it up for the very first time by a video guide (bottom row) automatically extracted from previous expert users. The time it took this novice user to do the task was 36s by simply re-watching the video guide twice.}
\label{fig:GlaciARExample}
\end{figure*}

Within the mixed reality continuum  that goes from the virtual to the real \cite{milgram94}, one of the arguably most noble applications of systems that augment us is the possibility to gain extra skills and guidance on how to do something either better or for the first time.

The ability to feel able to go about and do anything while being supported by an assistive  system has long been promised yet remains elusive. 
Consider being able to adjust your particular bike's chain by easy to follow guidance; receiving training in front of a new machine that others (but not you) have repaired before; showing up in any previously unknown kitchen anywhere, and seamlessly being shown where the key utensils are. All simple and common instances of where widely available guidance can help.

Despite the recent introduction of mass produced head-worn hardware systems such as those that are see through and that possess accurate visual positioning in space, and which appear to offer a way to get closer to better guiding systems, they have also put significant focus on an often overlooked aspect for mixed guidance systems: {\it the authoring problem}.

From the earliest systems such as Karma \cite{Karma}, guidance has assumed that 3D information is essential. But the reality is that 3D annotation is extremely hard to author and is so far only provided for well scripted tasks and results in this direction have always been wanting. Hardware limitations have in the past often been blamed but the authoring problem is, we argue, a much more serious obstacle which results in objects, places and sequences of interactions having to be mostly known in advance. The authoring bottleneck can be relieved somehow with in-situ annotation creation but to truly scale up the collection of key and nuanced information for any task, object, place and time, the spending of extra user time to author guides remains unconvincing.

On the other side, people is remarkably capable of following instructions as long as these are presented in a clear and intuitive manner and delivering guidance information with the wrong type of visualization or with overly synthesized information can be confusing at best and undermining the user's self confidence at worst. It is thus that the question of how best to augment a user's task guidance remains important.

In this work we are concerned with making some inroads into some of the above questions, namely 1) how to extract information in an unsupervised way for unscripted tasks by simply observing users so that mixed reality systems can start to scale up,  2) how to integrate a fully operational demonstrator of these ideas and 3) evaluate how the system actually supports people performing tasks.

The paper is organized as follows. In Section 2 we discuss guidance and authoring in MR/AR with emphasis on self-contained systems. Section 3 describes our approach starting with the description of the model of attention we use. Section 4 discusses the combination of the attention model with the video guide editor and object detector before in Section 5 we evaluate various aspects of the attention model performance including object discovery and multi user consistency. In Section 6 we evaluate GlaciAR with novice volunteers on three tasks before our discussion and conclusions.

\section{Guidance and Authoring}

We note that while there has been a large body of work on developing AR/MR systems for guidance, the vast majority of systems employ as part of the workflow pipeline an offline and supervised step for the authoring of the information to be displayed. This authoring can be from the earliest methods, text-based notes that appear when e.g. a tag appears in view \cite{Rekimoto} to more intricate 3D models that are meant to show assembly or repair instructions \cite{Henderson2011}. However as mentioned already, the authoring of the information to be displayed is non-trivial especially when we want to be able to perform guidance for any object or process anywhere. 

Making advances in ways in which guidance can be provided and the authoring bottleneck mitigated, will have a significant effect in augmenting users via MR/AR guidance.

A related system to ours is the Gabriel system \cite{Gabriel2015} which also uses Glass to guide users on tasks. In that case tasks are 2D such as assembling figures with colored blocks or drawing. That system implements a standard guidance and monitoring pipeline where the task is pre-scripted offline, as well as it uses an offboard strategy for information comparison and storage.

The system in \cite{Gabriel2015} implements verification of the manually scripted stages and does this by using relatively well constrained image metrics. In GlaciAR we do not enforce verification of a stage but rather concentrate in the arguably harder problem of automated capturing and delivery of information to inform the user about the task.

Other recent work has started to highlight the importance of automatically capturing workflows and the relevance of scaling up the authoring problem. In \cite{Petersen2012, Petersen2013}, a method that uses image similarities captures and monitors the workflow process of a task. The approach there compares incoming images with those stored for the same task and overlays them on a HMD. These overlays are a combination of mostly automated but also some manually authored information for the next step to follow. In \cite{cognitoPLOSOne}, the  workflow is captured by a rich array of on-body sensors and a semantic modeling of the task is used to keep track of the workflow. These approaches aim to alleviate the authoring problem somehow by trying to automate the extraction of relevant information as much as possible. In the case of \cite{Petersen2012, Petersen2013} this is done using 2D image metrics and the information to be displayed to users is an overlay on top of the object being part of the task. 

On the other hand, some concepts such as {\it Indirect Augmented Reality} \cite{Wither2011} have started to explore alternative MR/AR ways in which information can be delivered yet with less reliance on external positioning and associated hardware requirements.

Our work can be seen as effort in the above directions, but here we aim to push further the approaches of what information to capture and how to display it, both in fully automatic ways. Furthermore, most of the works on AR guidance assume that overlaying of information is crucial for guidance, but often this information can either be hard to understand, jittery due to 6D positional inaccuracies or at least substantially invading the visual field of view of the user. Our approach exploits the apparent limitations of contemporary eyewear computers e.g. Google Glass with a limited field of view and side-located display to provide short video snippets that have been automatically extracted from other users that have completed the task. In contrast to most other wearable guidance systems, we are aiming for extracting and presenting minimally distracting yet useful task information.

Previous work have developed methods to extract relevant information from a variety of sources that include eye tracking, 3D mapping and positioning via SLAM and visual object appearance detection \cite{youdoilearn2014}. But such approaches are not demonstrated in real-time and on eyewear hardware and assume that many sources of information are possible to collect. With GlaciAR, we explore a much more distilled and condensed authoring concept yet one that is amenable to be fully implemented onboard an eyewear system (Google Glass), in real-time and which allows for practical evaluation. The method we follow is illustrated in figure \ref{fig:GlaciARExample}  and described in the next sections.

\section{Method and Implementation}

\begin{figure}[t]
\centering
\includegraphics[width=0.9\columnwidth]{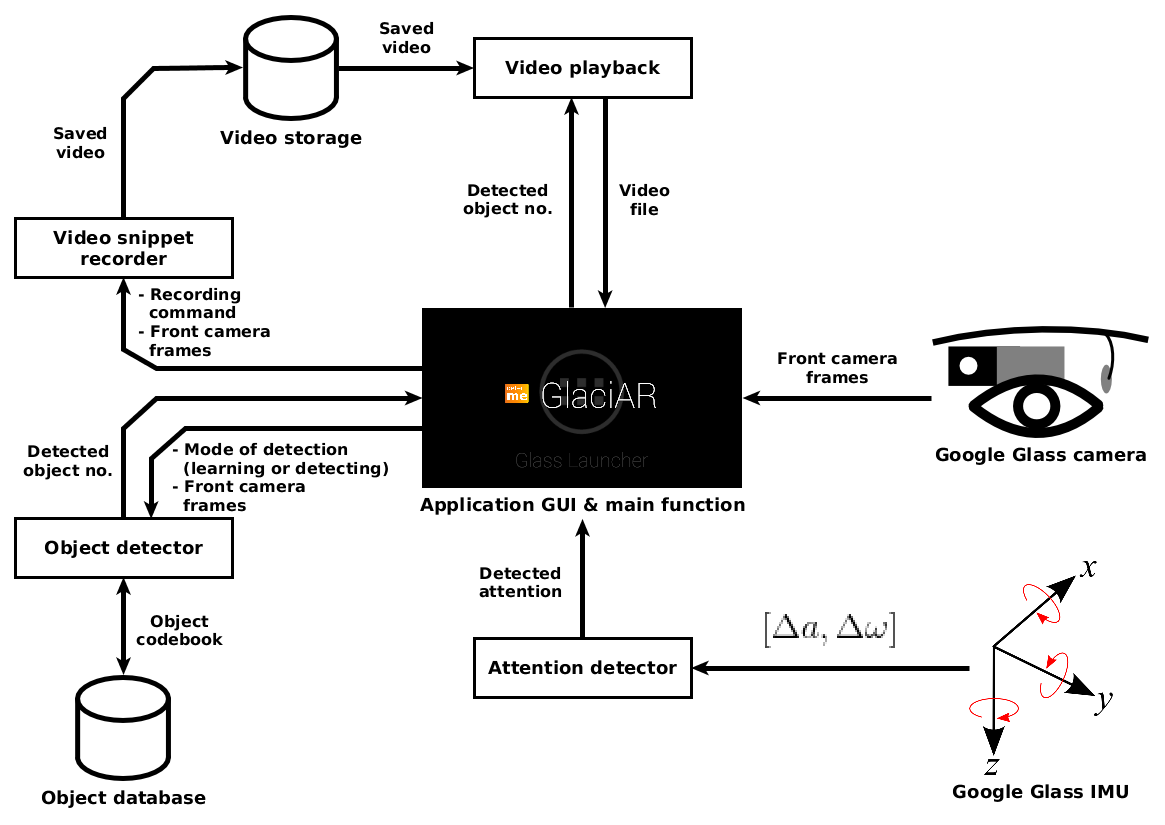}
\caption{For GlaciAR three key components are used: an attention detection module, a video snippet recorder which provides the instances of relevance and an object detector that triggers guidance.}
\label{fig:OverviewGlass}
\end{figure}

GlaciAR is underpinned by three interlinked components: 

\begin{itemize}
\item A model of user attention.
\item The capture of video snippets around instances of attention.
\item The detection of previously attended objects.
\end{itemize}

The overview of the system is shown in Figure~\ref{fig:OverviewGlass}. Two main sensors, the front facing camera and the inertial measurement unit feed information to the system. In GlaciAR, the module for attention determination is crucial since this is the mechanism used to make decisions of when to record or display information.

\subsection{Attention detection using Glass}

In other systems, eye-gaze has been useful as a source of attention determination via eye fixations \cite{youdoilearn2014}. That is, an angular velocity model for gaze fixations dictated when and where the person was paying attention. However in Glass there is no gaze tracker. We are thus inspired by the work in \cite{TeesidISWC2015} that shows the high level of correlation that exists between eye-gaze and head-gaze and the possibility to estimate spatial and temporal attention from the onboard IMU unit. For completeness, we briefly describe the approach followed and we substantially expand on the experimental evaluation of this attention model and evaluate its usefulness within the overall guidance system.

Important is to highlight that we are interested in moments of attention where the user is about to or already interacting with something. This is therefore a subset of all potential moments of attention that a user may have, yet by defining our attention for the instances of object interactions we aim to cater for an important set of the moments when the user is doing something of relevance and or needs guidance. We thus define the head-motion attention in a similar way as attention is often defined in gaze tracking, that is, a threshold to determine an eye fixation based on eye angular velocity \cite{Salvucci},  is here replaced by a threshold on the angular acceleration and velocities of head motion. We define temporal attention as

\begin{equation}
\textup{T}_{\textup{attention}} = \begin{cases}
\textit{attending,} & \text{ if } a \leqslant \tau ~ \text{ and } ~
\omega \leqslant \nu\\
\textit{in motion,} & \text{ Otherwise.}
\end{cases}, \label{eq:Moving}
\end{equation}

Where $\tau$ is the relative head acceleration threshold and $\nu$ is the relative head angular velocity threshold for identifying whether the user is attending to something or not. But this is only the temporal attention model, i.e. the {\it when} the user is paying attention. 

\begin{figure}[t]
\centering
\includegraphics[width=0.6\columnwidth]{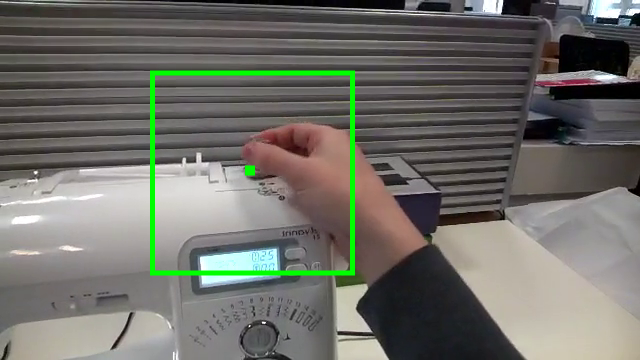}
\caption{Spatial attention position (green point) and the area of
interest (green box) acquired from Google Glass.}
\label{fig:GlassSpatialAttention}
\end{figure}

For spatial attention i.e. the {\it where} the user is looking at we opt for GlaciAR to use a fixed image location. This is backed by recent work that has investigated gaze fixations and that shows that for egocentric perception, when the user is interacting with things in the world, the location of where the user is fixating is concentrated around a small region in the image \cite{Fathi2013,TeesidISWC2015}. To compute the centre of mass of egocentric gaze fixations on the Glass' front camera image, we attach an eye gaze tracker to Glass and calibrate the location of where gaze is into it. With this information we can compute the gaze centre of mass for a number of people and tasks. We use images captured at 640x360 pixels instead of full resolution in order to reduce computational burden later and on these images the spatial attention point we use is located at coordinate $(250,189.5)$. The location toward the left side of the image is due to Glass' camera being mounted on the right side of the head. Figure \ref{fig:GlassSpatialAttention} shows the location of the fixed spatial attention coordinate. Note that the spatial attention region is only computed if the system is in the {\it attending} mode as per equation \ref{eq:Moving}.

GlaciAR's attention model is simple yet robust, requiring minimal computational burden and no image measurements. Extending the work of \cite{TeesidISWC2015}, in this paper we perform a more exhaustive evaluation on how useful this attention model is this to automated capture of relevant information.

\begin{figure}[h!]
\centering
{
  \includegraphics[width=0.95\columnwidth]{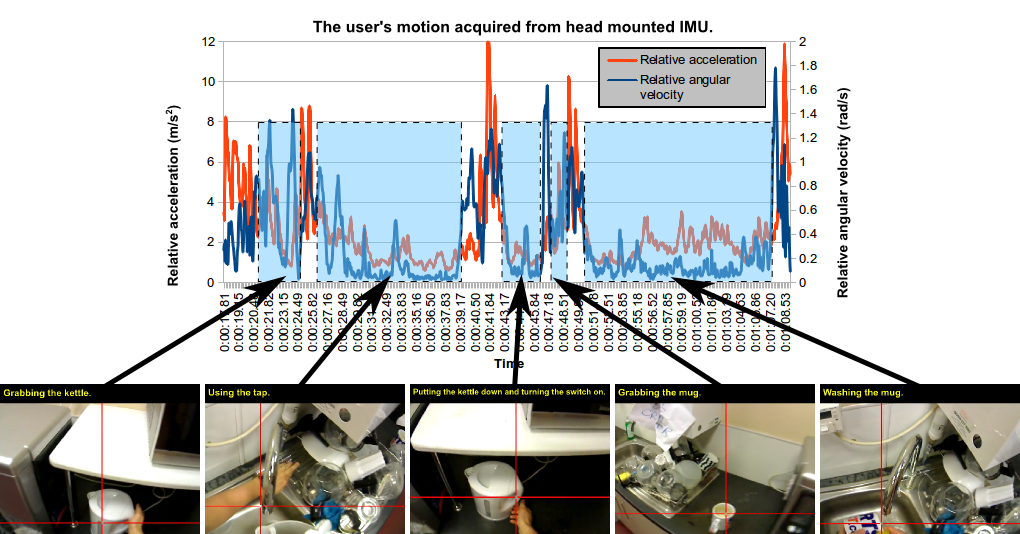}
}
\caption{The IMU signal acquired from Google Glass
during tasks performing of a user in a real environment setup.}
\label{fig:AttentionMotionA}
\end{figure}
 
\begin{figure}[h!]
\centering
{
  \includegraphics[width=0.95\columnwidth]{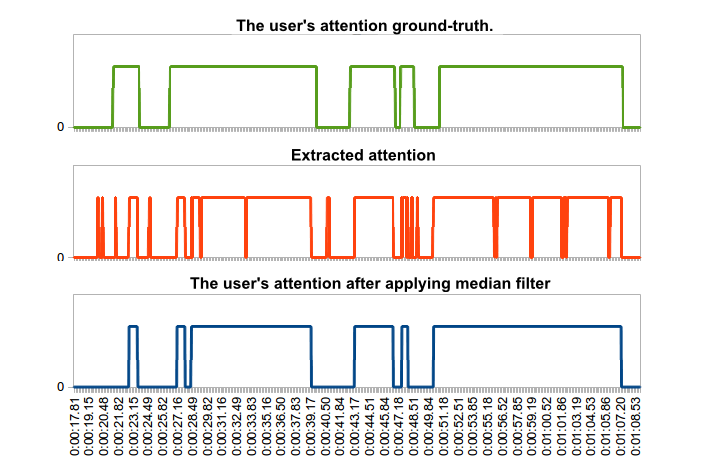}
}
\caption{The user's attention over time extracted from the IMU signal (middle)
using the threshold values $\tau=3.0~m/s^2$ and $\nu=0.5~rad/s$
compared to the ground-truth (top), and the user's attention after
applying a median filter (bottom).}
\label{fig:AttentionMotionB}
\end{figure}

Figure~\ref{fig:AttentionMotionA} presents the user's motion signals acquired from the Glass' IMU including the relative acceleration (red) and relative angular velocity (blue) as the user is making a cup of tea. The periods of time when the user is paying attention at the tasks are highlighted as the cyan-shaded rectangles and correspond to moments of eye fixation. The detected user's attention including manually selected ground-truth are shown in Figure~\ref{fig:AttentionMotionB}. In this figure, the extracted attention using the optimal threshold values as $\tau=3.0~m/s^2$ and $\nu=0.5~rad/s$ is shown in red. With these parameters, some of the actions on the ground-truth are classified as multiple independent actions rather than one action. In addition, some of the unattended events are also incorrectly identified as attention moments.

As a user is performing a task or moving around, there are chances that rapid movement or brief pauses happen. One way to mitigate the false positives is to filter the response. We use a median filter with a window size of 5 image frames which results in the smoothed signal that better corresponds with the ground truth (Figure~\ref{fig:AttentionMotionB} blue).

A more extended evaluation of the attention model for the specific task of extracting relevant and attended objects will be presented in Section 5.

\section{Automatic guide editing and linking with an object detector}

In GlaciAR, the attention model is the one that determines the when the person is doing something of relevance. This is based on the model described on the above section and is in contrast to other egocentric systems that use for example hand detection \cite{Mayol2005,Fathi2013} to indicate that something important is happening.

GlaciAR takes this approach for two reasons: first is that if the attention is linked to hands, the computational complexity required for the assessment increases substantially as hand detection in the wild is not trivial, and then, importantly a multitude of hand-eye coordination studies (e.g. see \cite{Land}) have shown that fixations (via gaze attention) precedes action by a good number of milliseconds. In this work, we thus hypothesize and evaluate how well a model of attention based on head motion  can also be used to preempt interactions and thus use it for the task of extracting video snippets around moments of interest. 

The approach is therefore to use the attention model instead of hand detection or any other environmental property as the director for video editing --- video snippet extraction starts and stops automatically when the system enters and leaves the {\it attending} mode (eq \ref{eq:Moving}).

The above approach can and indeed results in a number of videos captured as the user goes about interacting with objects. But to our advantage, most interactions with daily-living objects such as coffee machines, microwaves, car starting and similar are primarily driven by a reduced number or mostly a single way in which they can be interacted with. These tasks can include multiple steps such as to use a microwave "this button" needs pressing to open it, "this button" needs pressing for selecting the power option and "this knob" turned to select the time. All of these steps can and should form part of a single video guide on how to use the microwave.

\begin{figure}
\centering
\includegraphics[width=0.30\columnwidth]{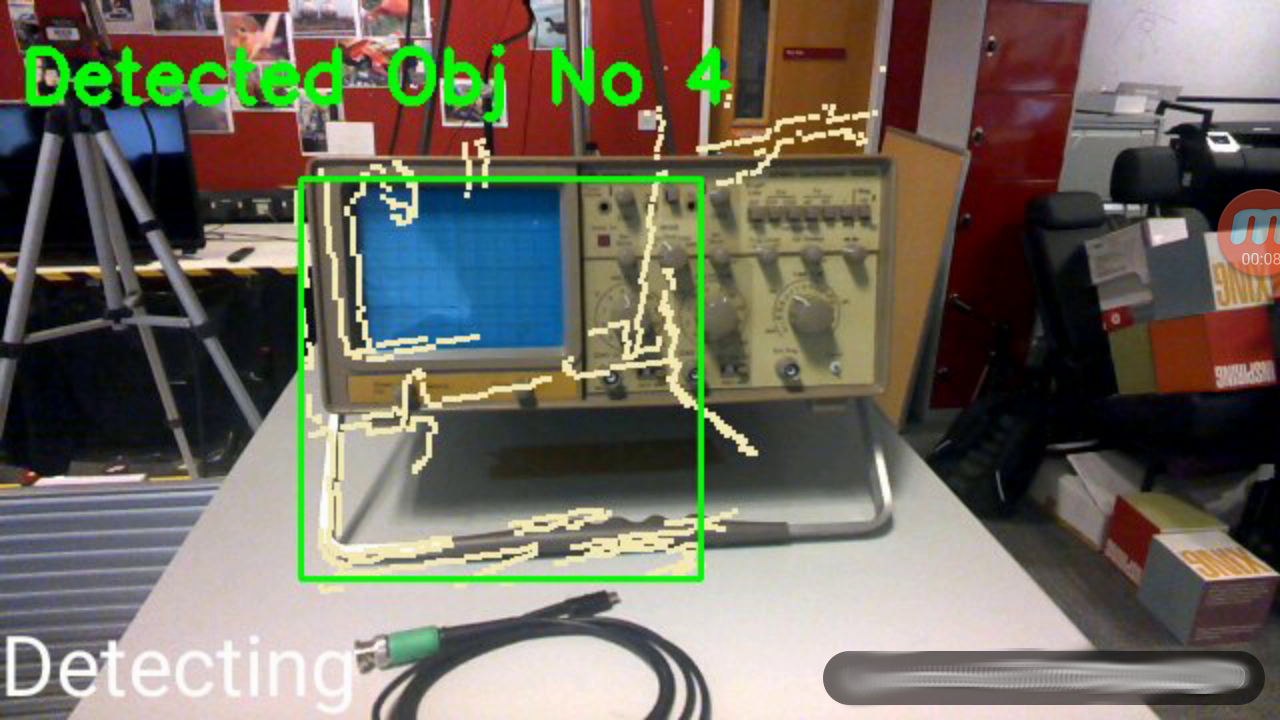}
\includegraphics[width=0.30\columnwidth]{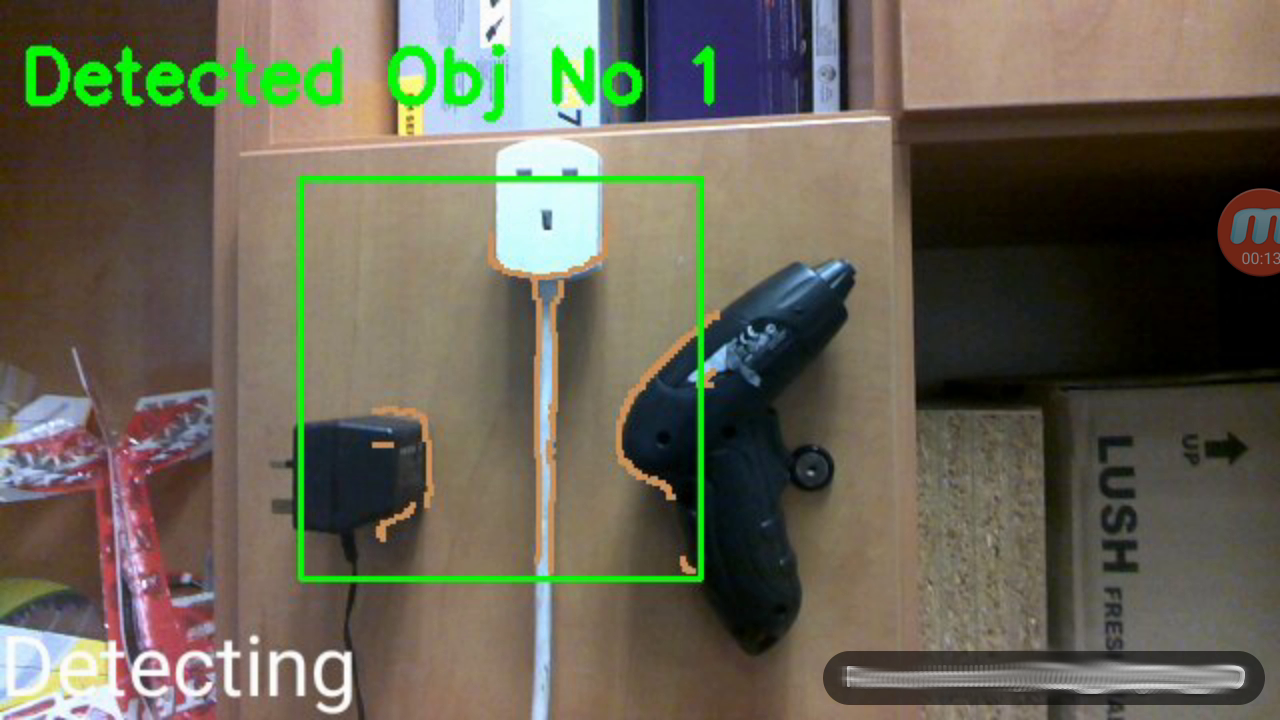}
\includegraphics[width=0.30\columnwidth]{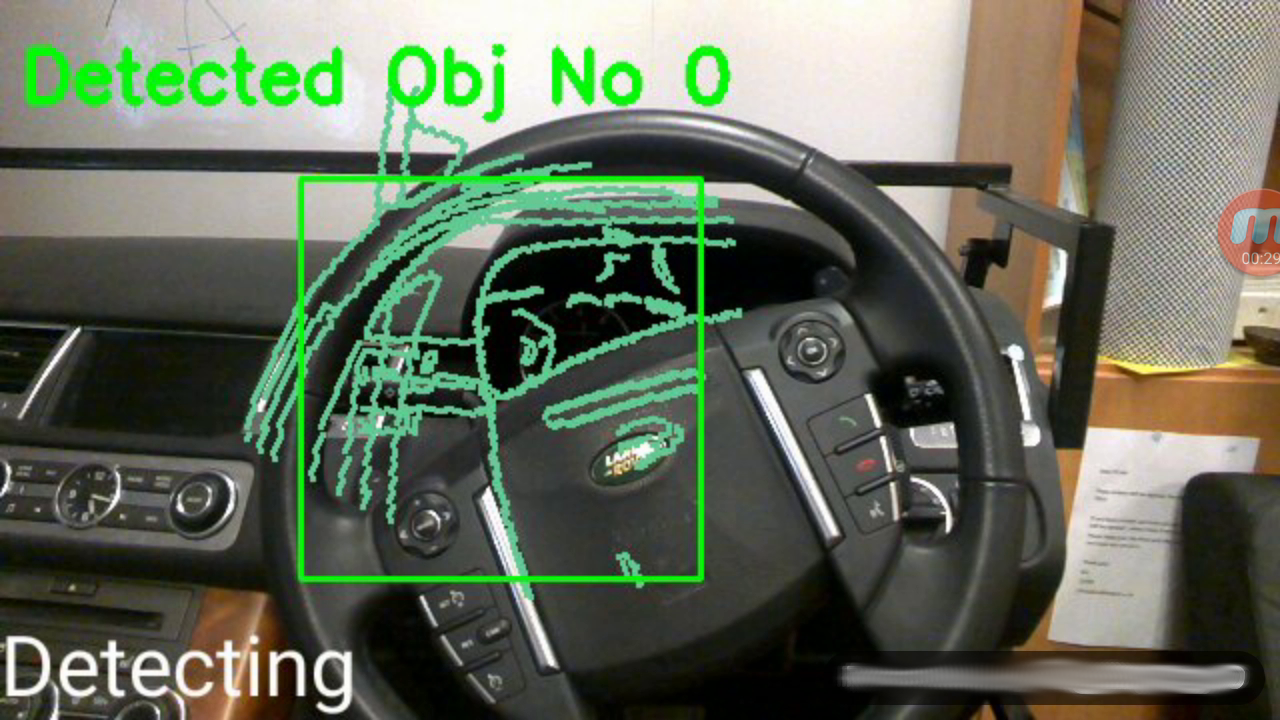}
\caption{The object detector runs onboard Glass and triggers video guidance when in front of an object previously used by experts. It uses the image only within the AOI estimated by the attention model. The overlaid edge indicates a successful detection.}
\label{fig:detections}
\end{figure}

The capturing of video snippets in the way GlaciAR works is not restrictive of collecting multiple ways in which the objects can be used. As per Figure \ref{fig:OverviewGlass} all collected videos are stored. For each one of these, the first frame for when the attention was detected and thus the object's {\it untouched} state is used to train a textureless object detector. For this an area of interest (AOI) of size 200x200 pixels (see figure \ref{fig:GlassSpatialAttention}) around the spatial centre of attention (250,189.5) is cropped and a descriptor based on edge configurations extracted \cite{detector2012}. This detector is computationally lightweight \cite{bunnunISMAR2012}, allows for multiple object detection, has invariance to scale, rotation and a degree of affine transformations and importantly runs entirely onboard Glass. All these contributes to reduced lag and robustness. 
In figure \ref{fig:detections}, example detections are shown. The detector runs from the information within the AOI estimated by the attention model which gates the image region and helps to keep computational demands low. 

When GlaciAR is in training mode, the model of attention captures videos linked to attention periods and trains the detector all in real-time. In this way it can simply observe expert users performing tasks while collecting the relevant information for guidance. When GlaciAR goes into assistive mode, the attention model is used to indicate that the user is interested in the object being attended, this prompts the detector to try to match the current AOI with stored ones and if a sufficiently good match is found, the associated video guide that was extracted from the expert's moment of attention is played on Glass. Note that since GlaciAR potentially captures one or more video guides for every expert in the training stage, the closest matching view when the novice is requiring guidance is displayed. This is the way in which novice users get guidance.

\section{Experimental evaluation}

We concentrate our evaluation first on the performance of the attention detection module since this is critical to all aspects in GlaciAR. One way in which we can evaluate its performance is via its ability to detect objects that the user has interacted with. To do this we follow the procedure used in \cite{TeesidISWC2015} where we attach a wearable gaze tracker (ASL Mobileye) to Glass for it to serve as one source of groundtruth. For this evaluation, we used 8 volunteers that were asked to wear the bundled device. This is a different group of users from the ones in the main results section. They were unaware of the type of information we were measuring so to reduce bias on the data collected. The volunteers were then asked to interact with objects around a building (e.g. open this door, press this button, lift that telephone, etc) guided by an investigator that used a stick with a coloured ``ping pong'' ball attached to it to indicate which object to use. This coloured taget was important to both unambiguously guide the interaction and discard imagery of any distracting saccades not part of the instructions, as well as to ensure there is ground truth since no current eye tracking hardware delivers results every time. Images from the scene facing camera of the gaze tracker were recorded at 30fps and synchronized with the IMU data on Glass narrowed to within a single image frame.

\subsection{Predicting interactions in advance}

As discussed before, eye gaze has been used widely to estimate attention and used as a precedent to action. In essence, people eye-gaze to what is about to be manipulated \cite{Land}. Since GlaciAR uses a head motion-based attention model it is important to ascertain how much predictive power it has and how much in advance it will be able to detect hand-object interactions compared to eye-gaze. Recall that if the interaction is predicted in advance a video snippet can then be captured shortly before this interaction takes place and the object detector will be trained without hands occluding it.
 
\begin{figure}[t]
\centering
	\includegraphics[width=1.0\columnwidth]{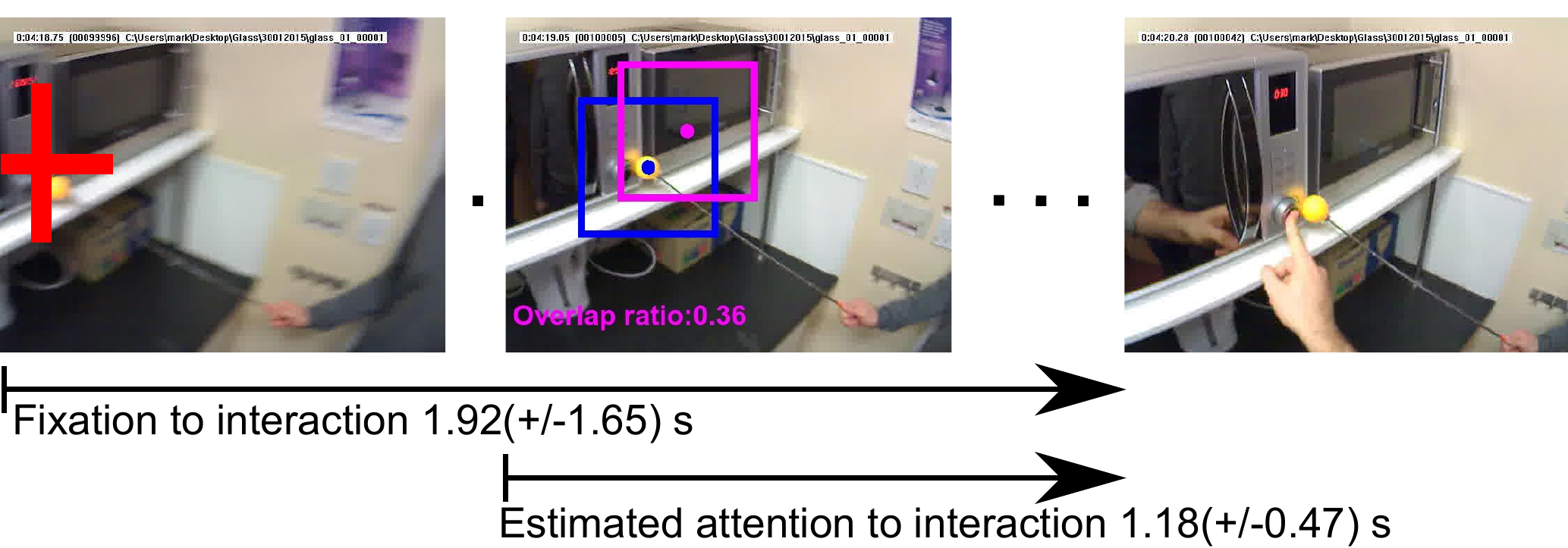}
\caption{Our method estimates hand-object interactions in advance. An
  object is gazed at by the user (red cross, left image), our method identifies
  user's attention (middle) before the user presses a button (right).}
\label{fig:AdvanceAttention}
\end{figure}

\begin{figure}
\centering{
\includegraphics[width=0.99\columnwidth]{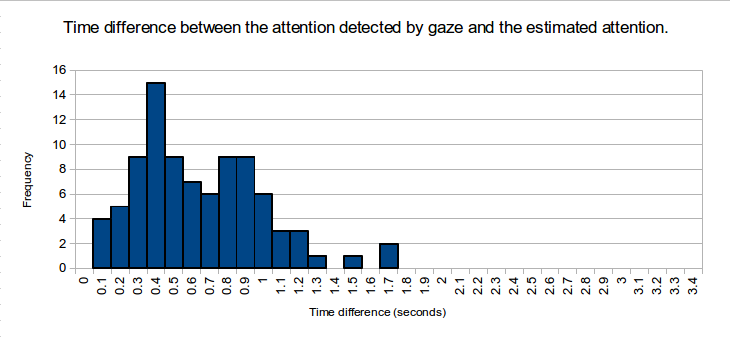}
}
\caption{The distribution of the difference between detecting attention with eye gaze and head motion. The attention is detected on average only about 0.6($\pm$0.35)s after that possible with gaze and overall 1.18($\pm$0.47)s in advance of a hand-object interaction.}
\label{fig:PredictionDistribution}
\end{figure}

Figure~\ref{fig:AdvanceAttention} illustrates the advance attention estimation. The figure shows an example of a user fixating at a microwave oven with eye-gaze, our head motion model then estimates attention and finally the microwave door is opened. Evaluating with a total of 91 hand-object of such interactions from the 8 users above mentioned, our approach can predict a hand-object interaction in advance on average 1.18($\pm$0.47) seconds before, and only after about 0.60($\pm$0.35) seconds of the attention estimated with gaze fixations (figure~\ref{fig:PredictionDistribution}). This ability of prediction based on the attention model, that results in a sufficient margin ($>$1000ms in advance), enables the operation of GlaciAR as described.

\subsection{Object discovery results}

Object discovery is a good measure to test how well the attention module is performing. We calculate object discovery via the intersection of two AOIs with size $200\times 200$ pixels one centered at the ground-truth (ping pong ball location) and the other centered at the attention estimation position. These two boxes were then compared using the standard PASCAL overlap criteria used for object discovery in Computer Vision, though note that no image processing is taking place here, all is driven by the IMU signal. We use a threshold of 30\% overlap between the AOIs to identify a discovered object. We then declared the object as true-positive discovery if the overlap is satisfied for 10 consecutive frames. This mitigates unstable and outlier discoveries.

The recordings used for this evaluation add to about 80 minutes of interactions (8 users x 10min/user).

\begin{figure}[t]
\centering
{
	\includegraphics[width=0.16\columnwidth]{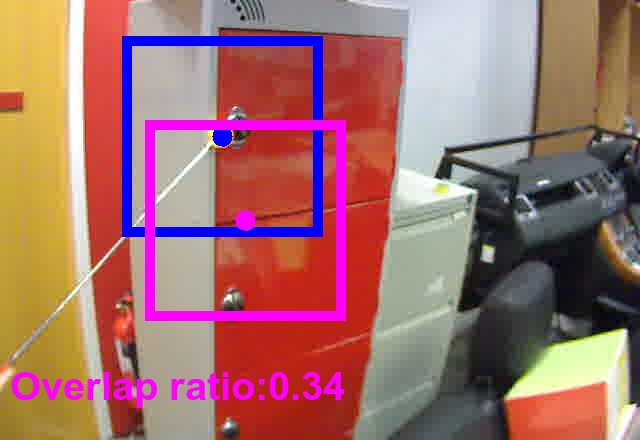}
	\includegraphics[width=0.16\columnwidth]{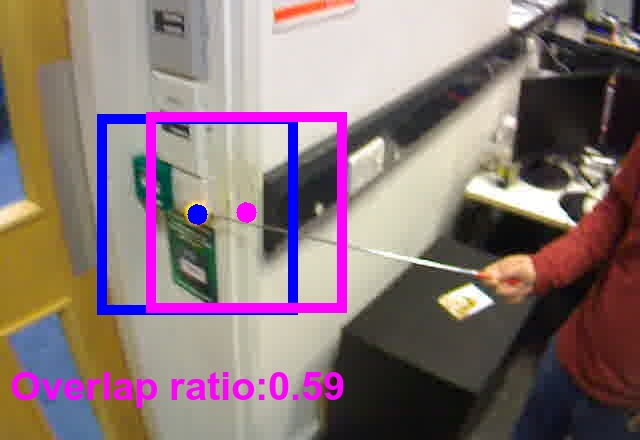}
	\includegraphics[width=0.16\columnwidth]{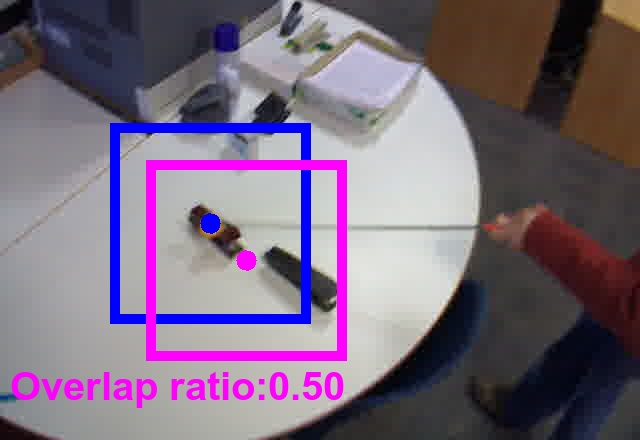}
	\includegraphics[width=0.16\columnwidth]{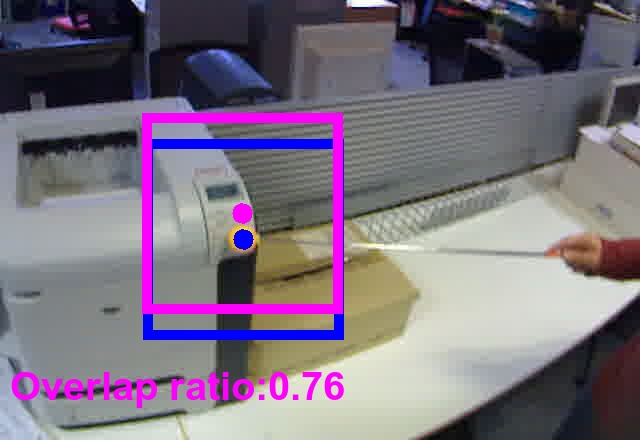}
	\includegraphics[width=0.16\columnwidth]{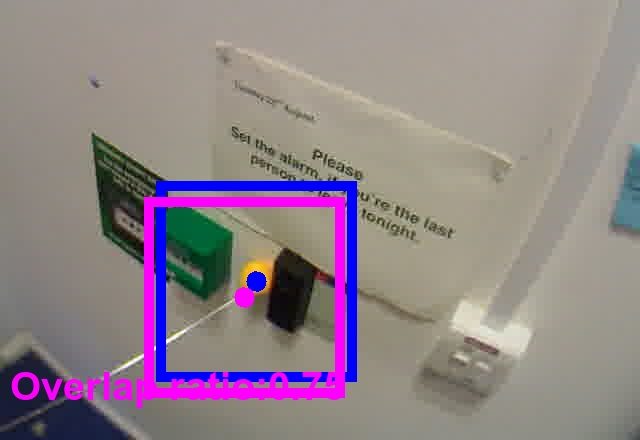} \\
	
	\includegraphics[width=0.16\columnwidth]{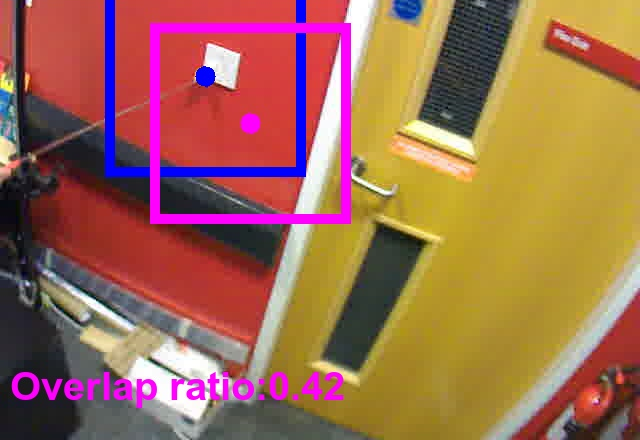}
	\includegraphics[width=0.16\columnwidth]{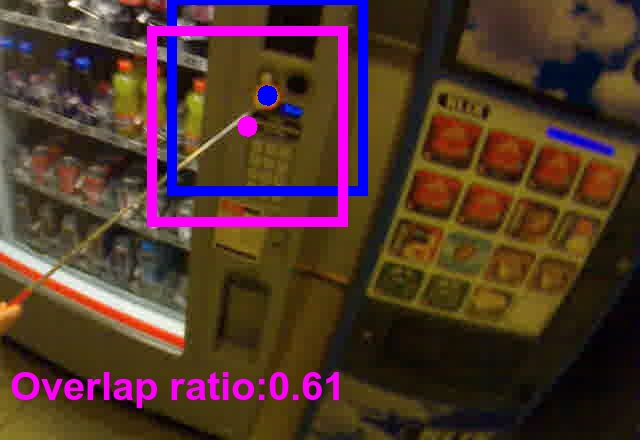}
	\includegraphics[width=0.16\columnwidth]{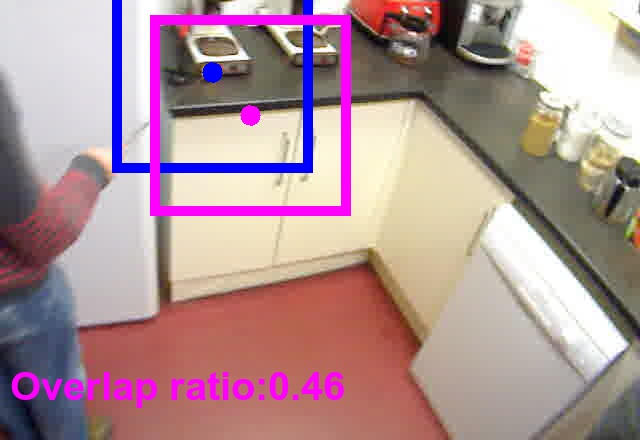}
	\includegraphics[width=0.16\columnwidth]{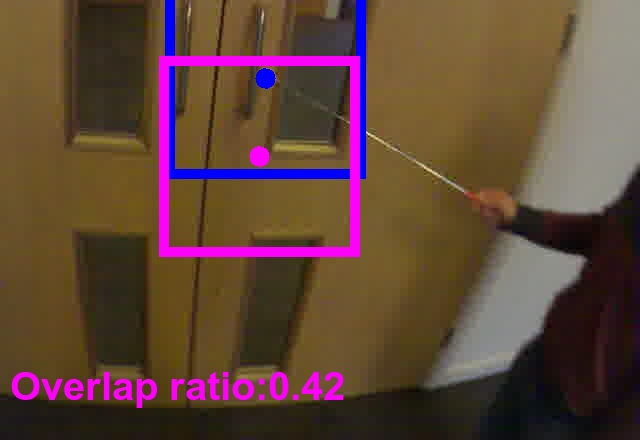}
	\includegraphics[width=0.16\columnwidth]{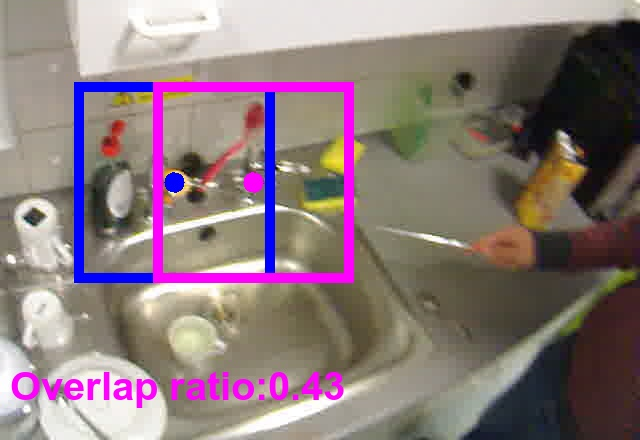} \\
	
	\includegraphics[width=0.16\columnwidth]{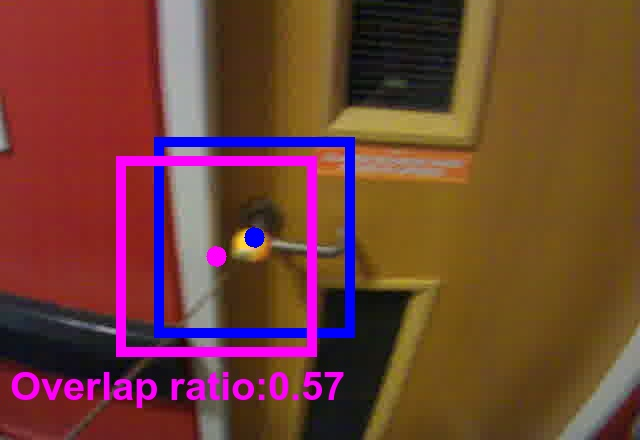}
	\includegraphics[width=0.16\columnwidth]{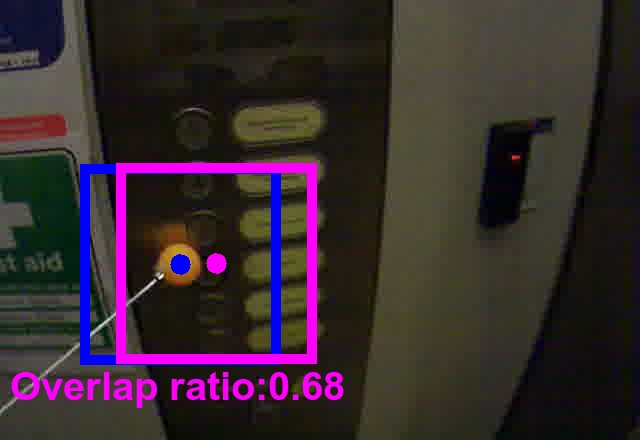}
	\includegraphics[width=0.16\columnwidth]{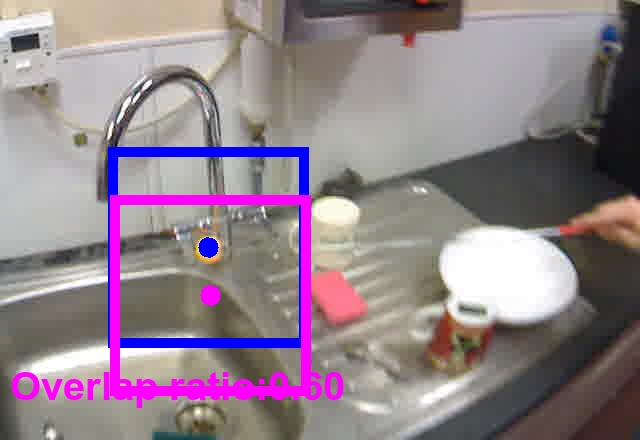}
	\includegraphics[width=0.16\columnwidth]{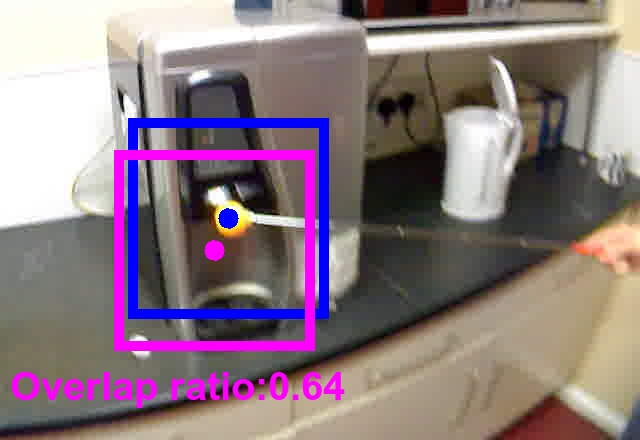}
	\includegraphics[width=0.16\columnwidth]{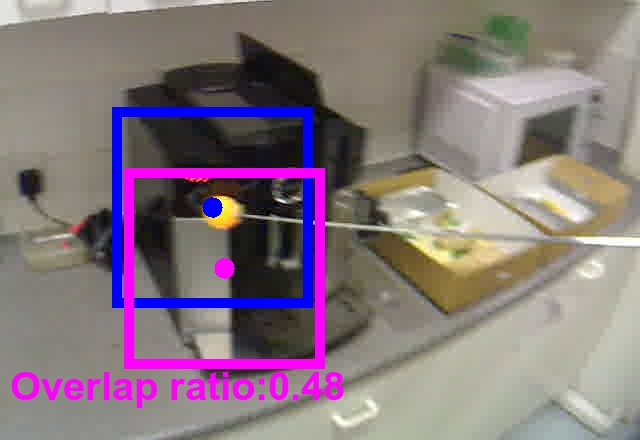} \\
	
	\includegraphics[width=0.16\columnwidth]{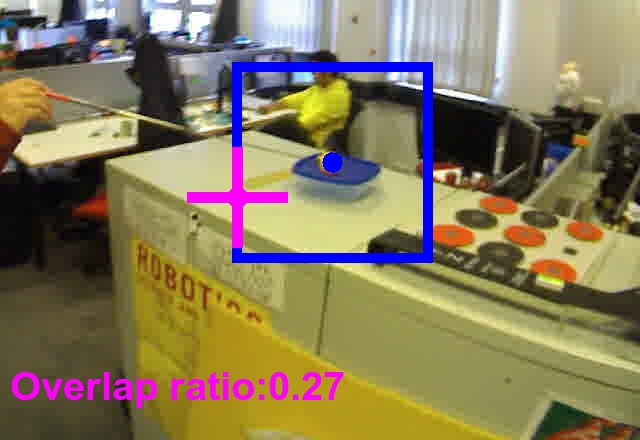}
	\includegraphics[width=0.16\columnwidth]{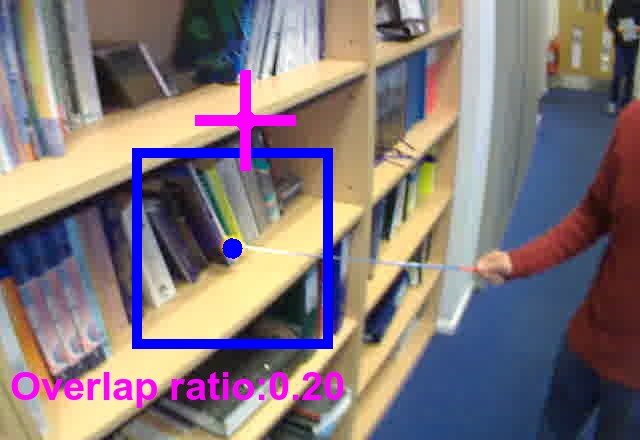}
	\includegraphics[width=0.16\columnwidth]{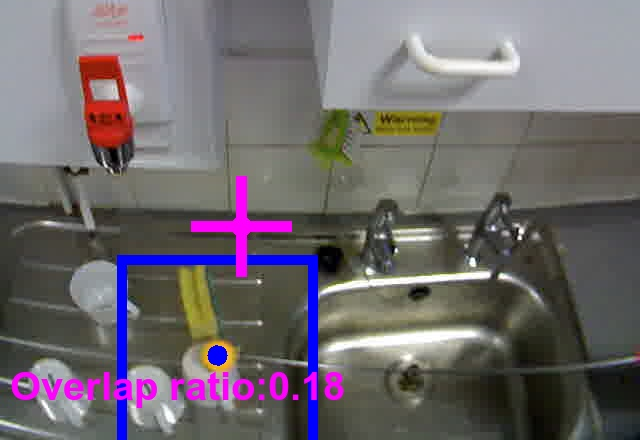}
	\includegraphics[width=0.16\columnwidth]{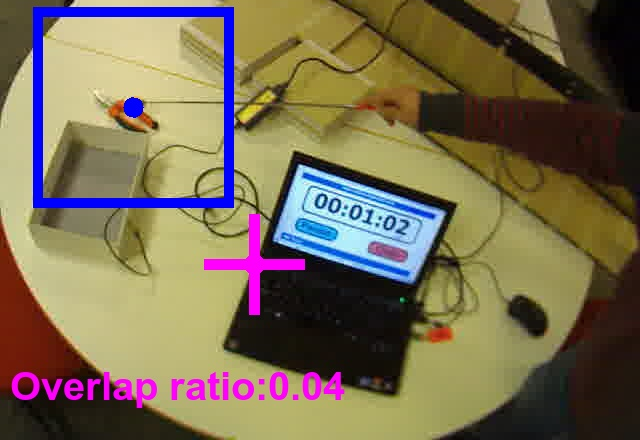}
	\includegraphics[width=0.16\columnwidth]{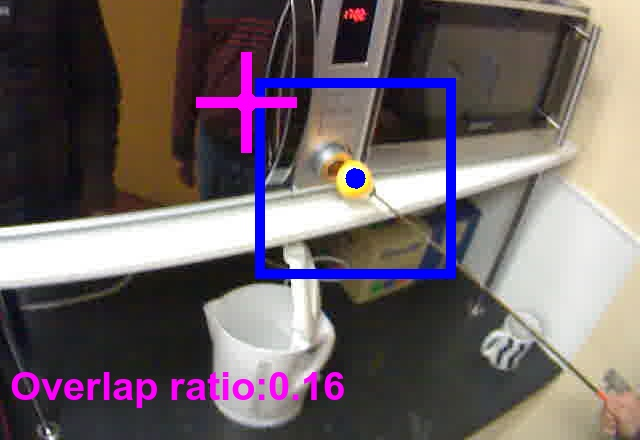}
}
\caption{Some results of object discovery from 3 different participants. Top 3 rows are example successfully discovered objects (magenta squares) and the bottom row shows missed discoveries, where the bounding box overlap criteria is not satisfied. But note most are just outside it. }
\label{fig:ObjectDiscovery}
\end{figure}

Examples of discovered and not discovered objects using the positions obtained from the spatial attention estimation are shown in Figure~\ref{fig:ObjectDiscovery}, where the ground-truth objects are within the blue coloured boxes, the successfully discovered objects are presented in magenta coloured boxes, and the missed discoveries  are marked as magenta crosses (bottom row in figure). As can be seen, the discovered objects correspond with objects interacted with, and that are within the overlap region. Several of the failed to be discovered objects are just outside the overlap criteria. In this test the object discovery precision is 0.61 and the recall 0.56. These results are conservative and a more relaxed AOI would result in increased rates, but at the expense of a larger image area to process in the detection stage.

\subsection{Multi-user object discovery}

For an assistive system that aims to learn from multiple people, another interesting question is what is the ratio of same-object discovery across multiple users as well as what type of object is more or less likely to be discovered. 

\begin{figure}
\centering{

\includegraphics[width=0.11\columnwidth]{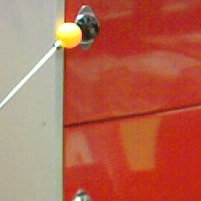}
\includegraphics[width=0.11\columnwidth]{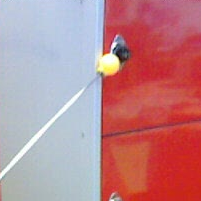}
\includegraphics[width=0.11\columnwidth]{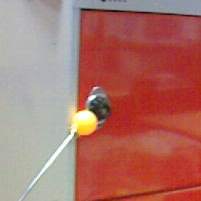}
\includegraphics[width=0.11\columnwidth]{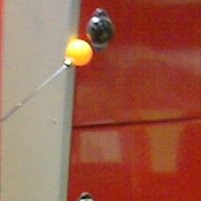}
\includegraphics[width=0.11\columnwidth]{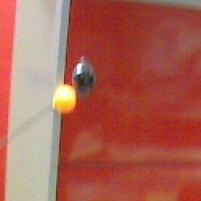}
\includegraphics[width=0.11\columnwidth]{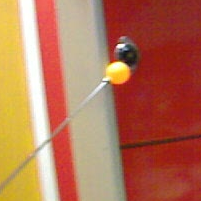}
\includegraphics[width=0.11\columnwidth]{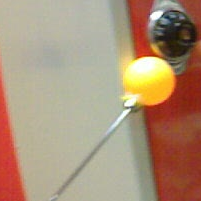}
\includegraphics[width=0.11\columnwidth]{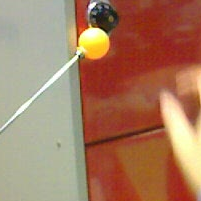}\\

\includegraphics[width=0.11\columnwidth]{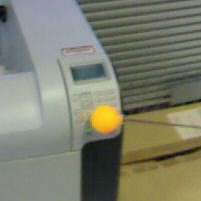}
\includegraphics[width=0.11\columnwidth]{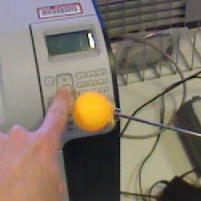}
\includegraphics[width=0.11\columnwidth]{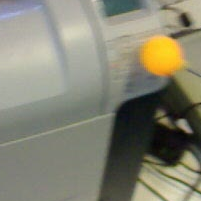}
\includegraphics[width=0.11\columnwidth]{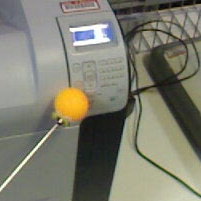}
\includegraphics[width=0.11\columnwidth]{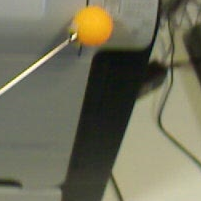}
\includegraphics[width=0.11\columnwidth]{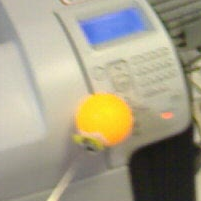}
\includegraphics[width=0.11\columnwidth]{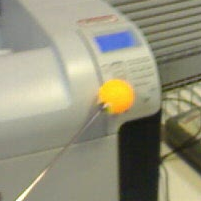}
\includegraphics[width=0.11\columnwidth]{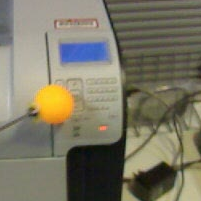}\\

\includegraphics[width=0.11\columnwidth]{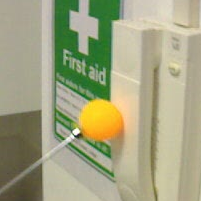}
\includegraphics[width=0.11\columnwidth]{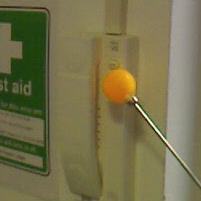}
\includegraphics[width=0.11\columnwidth]{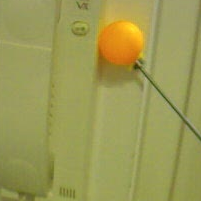}
\includegraphics[width=0.11\columnwidth]{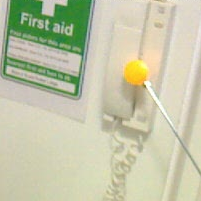}
\includegraphics[width=0.11\columnwidth]{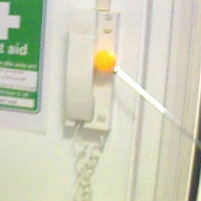}
\includegraphics[width=0.11\columnwidth]{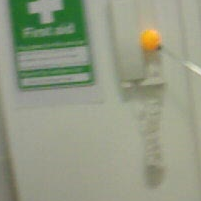}
\includegraphics[width=0.11\columnwidth]{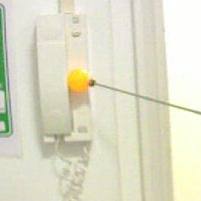}
\includegraphics[width=0.11\columnwidth]{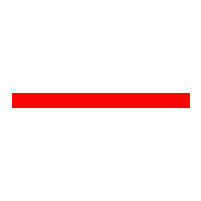}\\

\includegraphics[width=0.11\columnwidth]{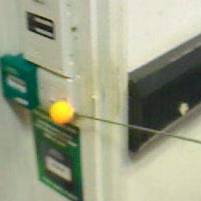}
\includegraphics[width=0.11\columnwidth]{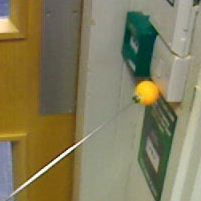}
\includegraphics[width=0.11\columnwidth]{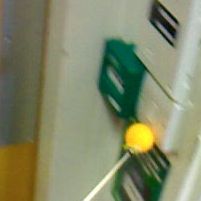}
\includegraphics[width=0.11\columnwidth]{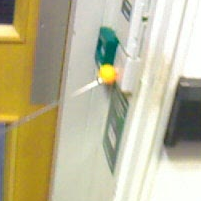}
\includegraphics[width=0.11\columnwidth]{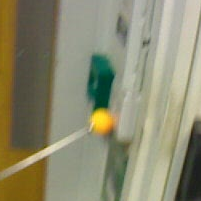}
\includegraphics[width=0.11\columnwidth]{NA.png} 
\includegraphics[width=0.11\columnwidth]{NA.png} 
\includegraphics[width=0.11\columnwidth]{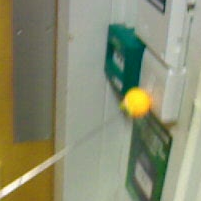}\\

\includegraphics[width=0.11\columnwidth]{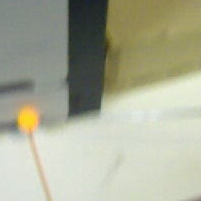}
\includegraphics[width=0.11\columnwidth]{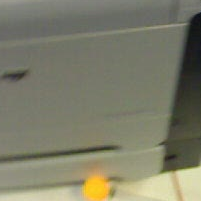}
\includegraphics[width=0.11\columnwidth]{NA.png}
\includegraphics[width=0.11\columnwidth]{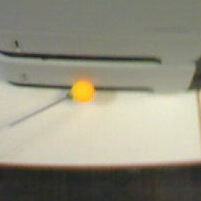}
\includegraphics[width=0.11\columnwidth]{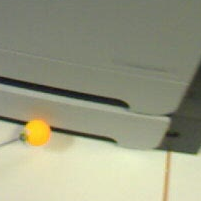}
\includegraphics[width=0.11\columnwidth]{NA.png}
\includegraphics[width=0.11\columnwidth]{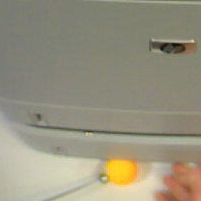}
\includegraphics[width=0.11\columnwidth]{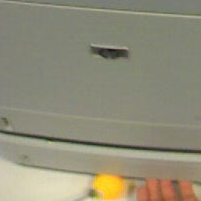}\\

\includegraphics[width=0.11\columnwidth]{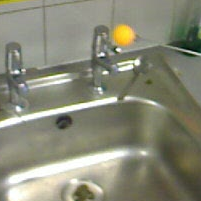}
\includegraphics[width=0.11\columnwidth]{NA.png}
\includegraphics[width=0.11\columnwidth]{NA.png}
\includegraphics[width=0.11\columnwidth]{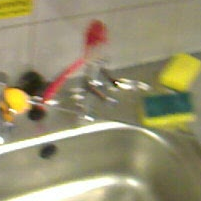}
\includegraphics[width=0.11\columnwidth]{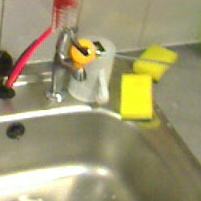}
\includegraphics[width=0.11\columnwidth]{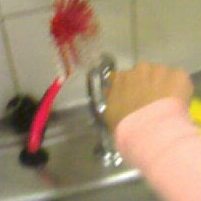}
\includegraphics[width=0.11\columnwidth]{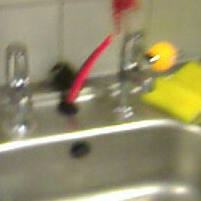}
\includegraphics[width=0.11\columnwidth]{NA.png}\\

\includegraphics[width=0.11\columnwidth]{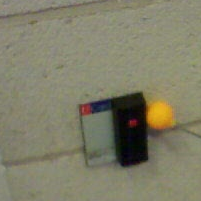}
\includegraphics[width=0.11\columnwidth]{NA.png}
\includegraphics[width=0.11\columnwidth]{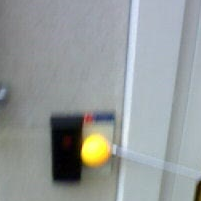}
\includegraphics[width=0.11\columnwidth]{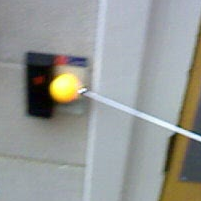}
\includegraphics[width=0.11\columnwidth]{NA.png}
\includegraphics[width=0.11\columnwidth]{NA.png}
\includegraphics[width=0.11\columnwidth]{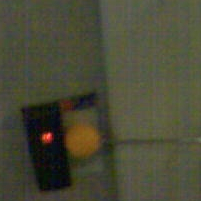}
\includegraphics[width=0.11\columnwidth]{NA.png} \\

\includegraphics[width=0.11\columnwidth]{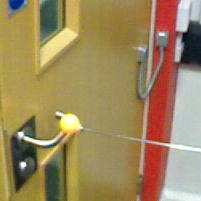}
\includegraphics[width=0.11\columnwidth]{NA.png}
\includegraphics[width=0.11\columnwidth]{NA.png}
\includegraphics[width=0.11\columnwidth]{NA.png}
\includegraphics[width=0.11\columnwidth]{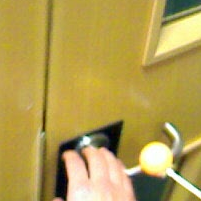}
\includegraphics[width=0.11\columnwidth]{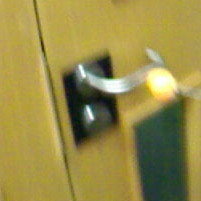}
\includegraphics[width=0.11\columnwidth]{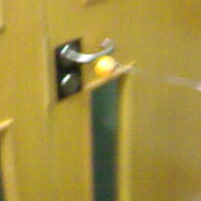}
\includegraphics[width=0.11\columnwidth]{NA.png}\\

\includegraphics[width=0.11\columnwidth]{NA.png}
\includegraphics[width=0.11\columnwidth]{NA.png}
\includegraphics[width=0.11\columnwidth]{NA.png}
\includegraphics[width=0.11\columnwidth]{NA.png}
\includegraphics[width=0.11\columnwidth]{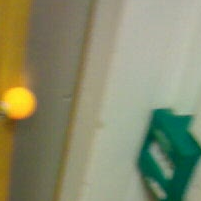}
\includegraphics[width=0.11\columnwidth]{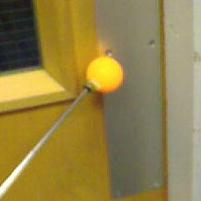}
\includegraphics[width=0.11\columnwidth]{NA.png}
\includegraphics[width=0.11\columnwidth]{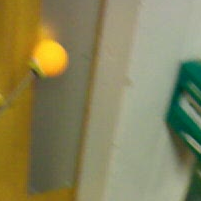}\\

\caption{Nine objects (rows) being discovered from 8 different users (columns) using our
attention method as users explore various objects. Objects that failed to be detected by a specific user are labeled as a red dash. The worst performing discoveries are for objects that are interacted with only briefly e.g. doors (last rows).}
\label{fig:CommonObjs}
}
\end{figure}

On our experimental sequences there were 9 common objects that all participants were asked to interact with. These are: a safe cabinet, a printer panel, a telephone, a door release, a printer tray, a tap, a card reader, a door handle and a door plate. These are shown in this order in figure \ref{fig:CommonObjs}. Per user, our method discovered 5.66($\pm$1.8) objects on average and achieved a recall rate of 0.71. Importantly, all objects were discovered by 3 or more users, which incidentally highlights the importance of multiple users collecting the information we require.

But some of the objects posed greater challenge. In particular, objects that are only very briefly interacted with under naturalistic conditions such as a door handle as it is opened, are the harder to be detected (bottom rows in figure \ref{fig:CommonObjs}), while those other objects that are operated in the style of a printer's panel, answering a telephone or opening a safe are more easily detected by all users (top rows in figure \ref{fig:CommonObjs}).

\section{Guidance Evaluation}

This section describes the evaluation of GlaciAR where tasks are
automatically captured, edited and delivered to guide users.
GlaciAR runs onboard a Google Glass Explorer Edition 2.0, such that
all attention detection, snippet video editing, object detection, and video
guide delivery are run in real-time and without user intervention. 

The way in which GlaciAR is triggered for guidance is as the user walks up to an object where it has previously been trained it detects such object and delivers the video guide closest to the detected viewpoint. We are in this work concerned with guidance evaluation and not the object detection performance which is subject to a separate analysis. There are thus two main tests conducted in the experiments. Firstly, performance evaluation of the attention model and then an evaluation of actual guidance performance.

\subsection{Video Guide Authoring Evaluation}

This section compares the effectiveness of GlaciAR in automatically extracting video guides 
versus video guides manually edited by expert users. This aims to demonstrate the extent of effectiveness of the attention-driven harvesting of information as well as the overall concept of video guidance from snippets. There are two conditions as follows.

\begin{itemize}
  \item Automatically generated video guide: the video guides are
  automatically extracted by GlaciAR using the proposed model of attention.
  \item Expert generated video guide: the video guides are manually
  edited by experts, who indicate the start and end of the activities.
\end{itemize}


Our hypothesis before conducting the evaluation is that if the attention-driven model is useful, there will be little to no statistical difference between the performance achieved with the expert edited video guides and the ones automatically extracted by GlaciAR.

We had two operational sessions, training and testing. Three experts
participated in the training session and 14 novice volunteers
(eight females, six males) aged from 24 to 36 were recruited. None of
the novice volunteers had participated in the our previous studies or possessed
any prior experience with Google Glass.

\begin{figure}[h!]
\centering
\subfloat[Oscilloscope]
{
\includegraphics[width=0.30\columnwidth]{o-scope.png}
\label{fig:Osc}
}
\subfloat[Screwdriver]
{
  \includegraphics[width=0.30\columnwidth]{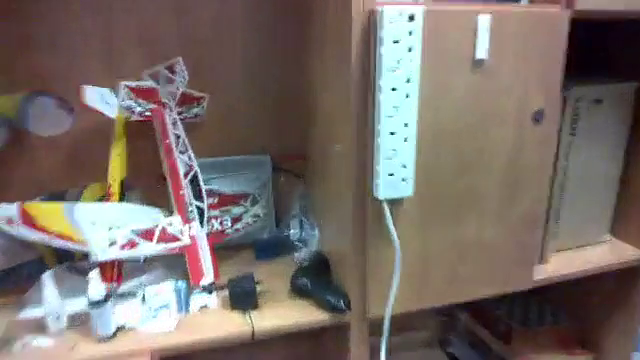}
  \label{fig:Screwdriver}
}
\subfloat[Sewing machine]
{
  \includegraphics[width=0.30\columnwidth]{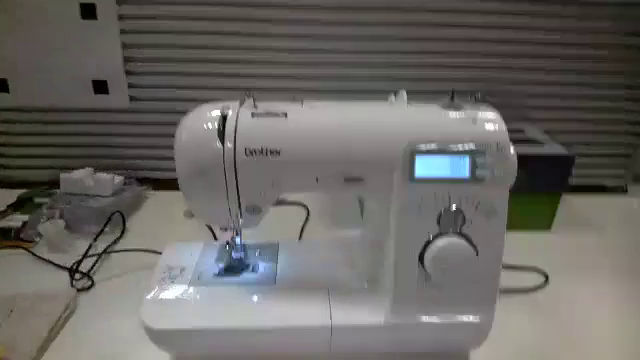}
  \label{fig:Sewingmachine}
}
\caption{The objects used in the video guide evaluation.}
\label{fig:ScrewdriverNSewingmachine}
\end{figure}

The tasks involved an oscilloscope, an electric screwdriver and a sewing machine
 (Figure~\ref{fig:ScrewdriverNSewingmachine}). Each one of these tasks have an increasing number of steps from three to five.
Each expert performed every task, and then was asked to watch and
edit a video guide of the task by indicating the start and end of the task. In total, there were 9
videos (3 videos $\times$ 3 tasks) for each video guide condition. The
task stages are  described in Table~\ref{tab:TaskExplanation}.

\begin{table}[h!]
\centering
\begin{tabularx}{\linewidth}{|l|X|}
   \hline
   \multicolumn{1}{|c|}{Task} & \multicolumn{1}{c|}{Process} \\
  \hline
  Oscilloscope (Osc) - 3 steps
   & \textbullet Attach the probe to the socket. \\
   & \textbullet Turn on the switch. \\
   & \textbullet Adjust the trace to zero. \\
  \hline
  Screwdriver (Scr) - 4 steps
   & \textbullet Pick up the adaptor and plug it into the socket.\\
   & \textbullet Pick up the screwdriver and attach the adaptor's \\
   & cord to the screwdriver. \\
   & \textbullet Put the screwdriver back on the shelf. \\
   & \textbullet Turn on the socket switch. \\
  \hline
  Sewing machine (Sew) - 5 steps
   & \textbullet Open the drawer. \\
   & \textbullet Pick up the bobbin and put it in the bobbin pin. \\
   & \textbullet Push the bobbin pin to the right to lock it. \\
   & \textbullet Pick up the purple spool and put it in the spool pin. \\
   & \textbullet Press the button to start spinning the bobbin. \\
  \hline
\end{tabularx}
\caption{Task descriptions in the video authoring evaluation.}
\label{tab:TaskExplanation}
\end{table}

During the testing session, each participant was asked to perform
every task and their performance was recorded for analysis. After each task, the participant was also asked to fill out a NASA-TLX survey and an opinion feedback questionnaire.
To be able to assess users close to our originally stated ambition of anywhere augmentation, the participants performed each task only once. Performing a task several times on the two different test conditions would have biased the
evaluation, as the participants would already have been aware of the task after the first trail.
The video guide condition, i.e. automatic or expert cut, was randomly
assigned and the participant was not aware of which condition was being delivered.
Each participant, therefore, performed all three tasks watching two video
guide conditions, i.e. two automatic cut videos and one expert cut
video, or one automatic cut video and two expert cut videos.
Hence, each video authoring condition was used 7 times in every task.

\subsubsection{Video Guide Authoring Results}

The automatic generated video guides were automatically extracted by GlaciAR while the experts performed the task during the training session. After that, the
experts were given their videos of entire activities in the training
session and asked to indicate the start and end point in each video.

\begin{figure}[!h]
\centering
{
\includegraphics[width=0.48\columnwidth]{mike-osc1.png}~
  \includegraphics[width=0.48\columnwidth]{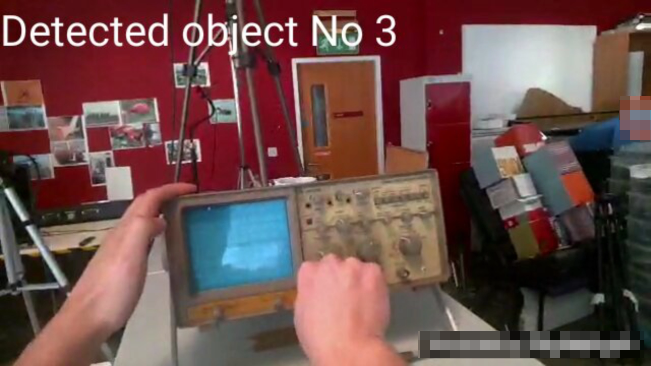}
  }\\
{
\includegraphics[width=0.48\columnwidth]{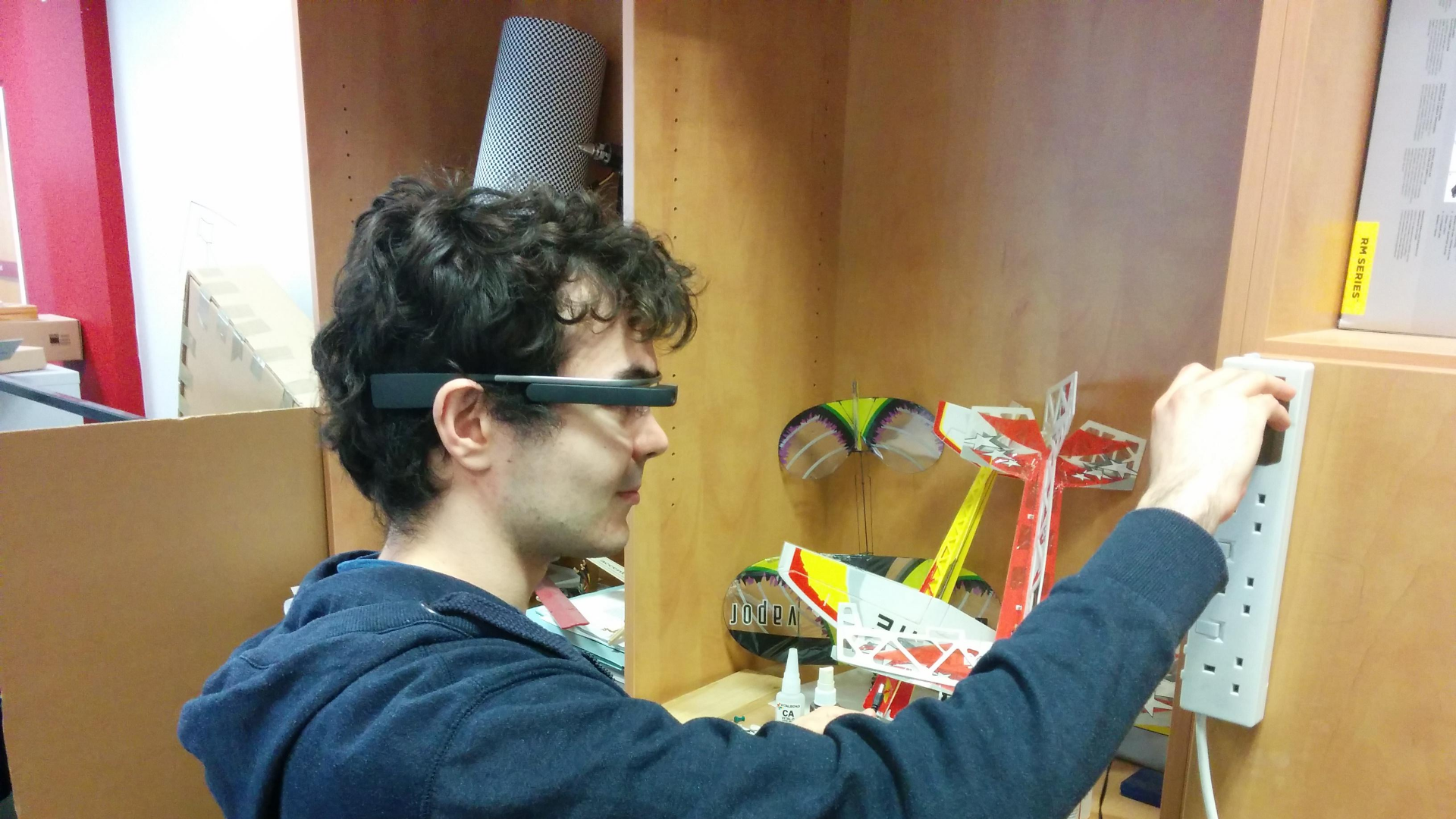}
\includegraphics[width=0.48\columnwidth]{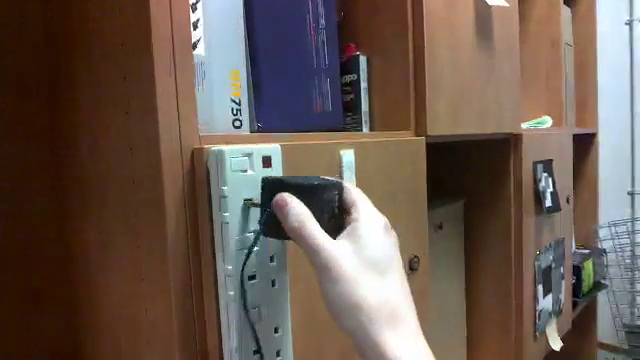}
}\\
{
\includegraphics[width=0.48\columnwidth]{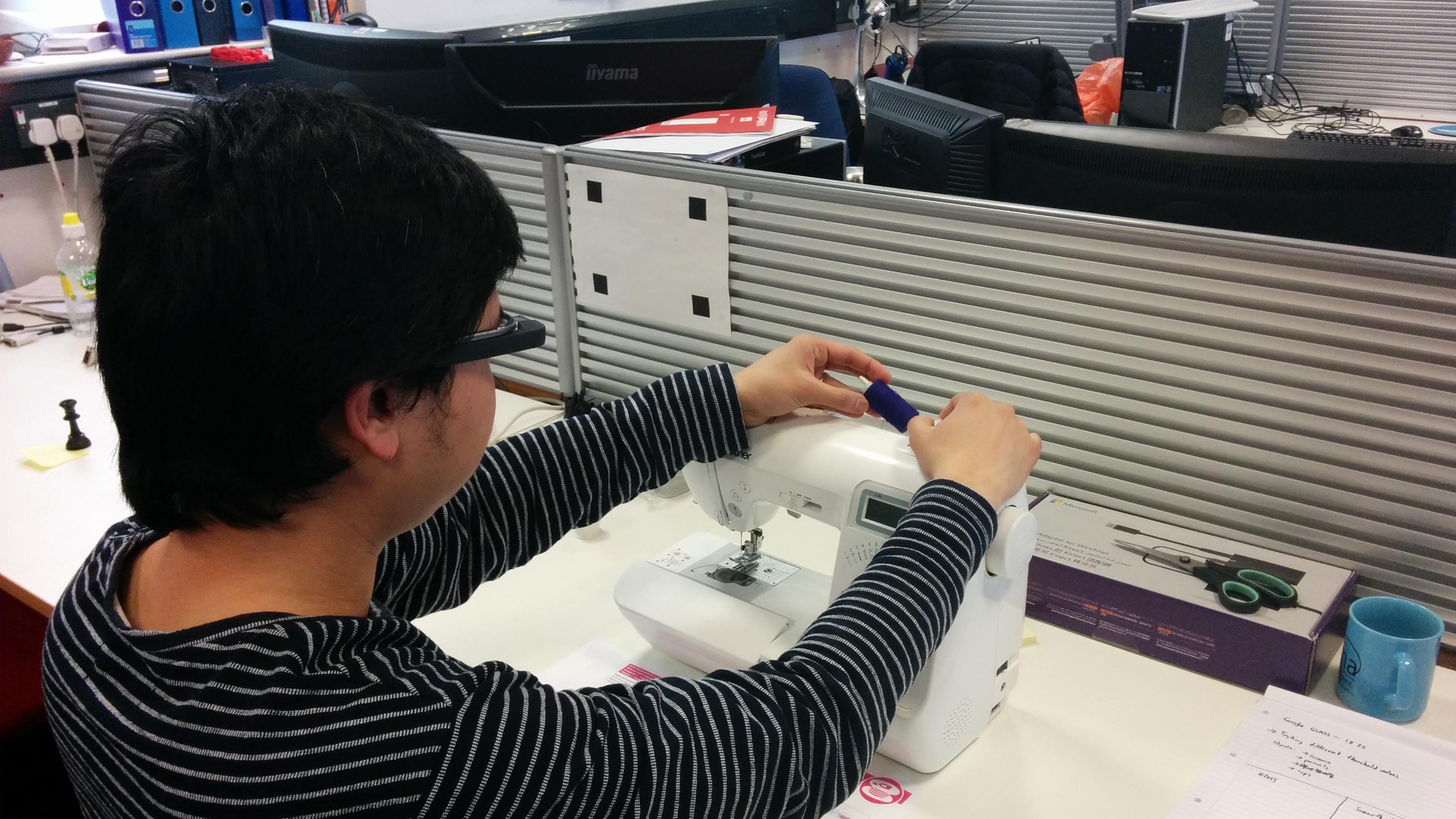}
\includegraphics[width=0.48\columnwidth]{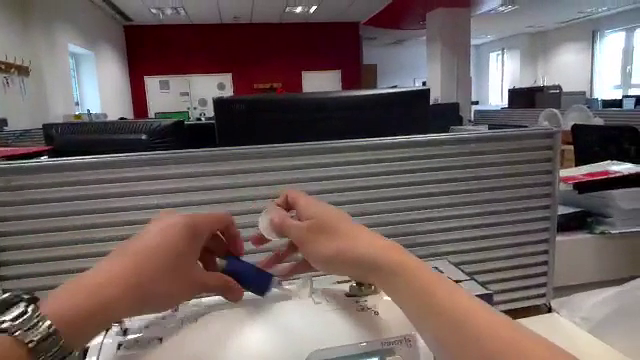}
}
\caption{Images showing the wearers using GlaciAR and its screenshots
in the three evaluated tasks.}
\label{fig:ScrewdriverSewing}
\end{figure}

\begin{table}[h!]
\centering
\begin{tabularx}{\columnwidth}{|l|l|c|X|}
\hline
\multicolumn{1}{|c|}{\multirow{2}{*}{Task}} &
\multicolumn{1}{c|}{Video} & Average video & Overlapping \\
& \multicolumn{1}{c|}{condition} & length (s) & percentage\\
\hline
\multirow{2}{*}{Oscilloscope} & Automatic & $16.3(\pm3.24)$ &
\multirow{2}{*}{89.71\%} \\
\cline{2-3}
 & Expert & $16(\pm2.65)$ & \\
\hline
\multirow{2}{*}{Screwdriver} & Automatic & $14.43(\pm3.20)$ &
\multirow{2}{*}{86.69\%} \\
\cline{2-3}
 & Expert & $16.33(\pm3.21)$ & \\
\hline
\multirow{2}{*}{Sewing machine} & Automatic & $21.87(\pm3.19)$ &
\multirow{2}{*}{85.97\%} \\
\cline{2-3}
 & Expert & $25.00(\pm3.60)$ & \\
\hline
\end{tabularx}
\caption{The average length (s) of the video guide generated by
GlaciAR and expert cut, and the percentage of overlapping (\%) of
the automatic and expert authoring video guide condition.}
\label{tab:VidAuthoringTime}
\end{table}

Figure \ref{fig:ScrewdriverSewing} shows the testing scenarios and the
screenshots taken from the Google Glass screen showing the video guide
instructions of the task. Table~\ref{tab:VidAuthoringTime} presents
the average length of the video guide extracted by both conditions in
each task. The average video length results of the automatic condition
are similar to the ones extracted using the experts' indications.
Furthermore, the results of the percentage of overlapping also show
that the scores are over 85\% in every task. This already suggests high
similarity between the automatic and the expert edits but still requires confirmation of performance to back the hypothesis. This is evaluated next.

\subsubsection{Success Rate and Completion Time}

The task's success is measured as a percentage of the task that a user can accomplish. For example, for five steps in the sewing machine task, if a user succeeds in four steps, that means the user has achieved 80\% of the process.

\begin{table}[h!]
\centering
\begin{tabular}{|l|c|c|c|c|}
  \hline
  \multicolumn{1}{|c|}{\multirow{2}{*}{Task}} &
  \multicolumn{1}{c|}{Video} & Success & Completion & No. of \\
   & \multicolumn{1}{c|}{condition} & rate & time & videos \\
  \hline
  \multirow{2}{*}{Osc} & Auto & 100 & $67.14(\pm9.67)$ &
  $4.00(\pm1.00)$
  \\
  \cline{2-5}
   & Expert & 100 & $67.00(\pm12.17)$ & $3.86(\pm0.38)$ \\
  \hline
  \multirow{2}{*}{Scr} & Auto & 92.86 & $73.29(\pm20.30)$ &
  $3.86(\pm0.69)$
  \\
  \cline{2-5}
   & Expert & 92.86 & $71.14(\pm22.26)$ & $4.00(\pm1.15)$ \\
  \hline
  \multirow{2}{*}{Sew} & Auto & 85.71 & $134.86(\pm48.22)$ 
  & $6.14(\pm1.95)$ \\
  \cline{2-5}
   & Expert & 85.71 & $123.00(\pm34.93)$ & $4.71(\pm0.76)$ \\
  \hline
\end{tabular}
\caption{Success rate (\%), average completion time
(seconds) and average number of videos (times) needed for performing the tasks.}
\label{tab:SuccessfulRateNCompletionTime}
\end{table}

\begin{figure}[h!]
\centering
\subfloat[Oscilloscope] {
  \includegraphics[width=0.32\columnwidth]{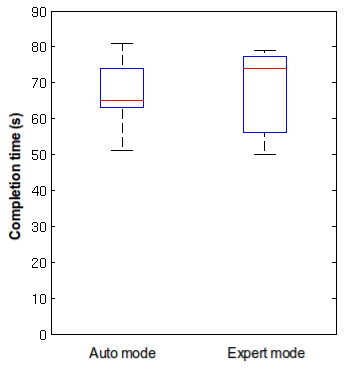}
  \label{fig:CompletionTimeA}
}
\subfloat[Screwdriver] {
  \includegraphics[width=0.32\columnwidth]{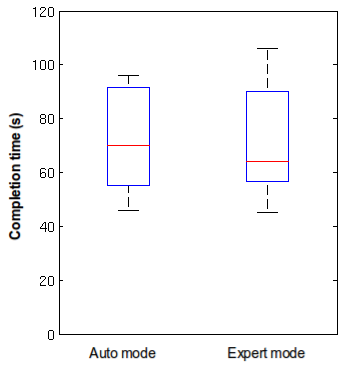}
  \label{fig:CompletionTimeB}
}
\subfloat[Sewing machine] {
  \includegraphics[width=0.32\columnwidth]{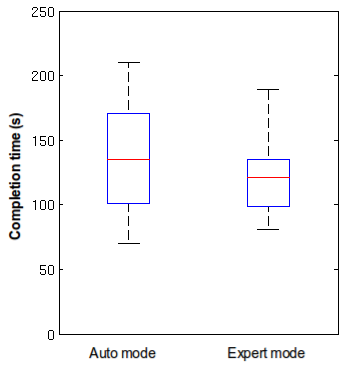}
  \label{fig:CompletionTimeC}
}
\caption{Boxplots illustrating the comparison of completion time of
each task between the automatic and expert video guide authoring
condition.}
\label{fig:CompletionTime}
\end{figure}

Table~\ref{tab:SuccessfulRateNCompletionTime} presents the tasks'
success rate, the average completion time, and the average number of
videos. The success rates are at 100\% for all in the Oscilloscope task in
both video authoring conditions. However, in the screwdriver
task, the success rates in both conditions are at 92.86\%. There were
four participants who reported that they could not see `the turn on
the socket switch' action at the end of the video guide due to the
expert's hand blocking the view. In the sewing machine task (85.71\%
successful rate), there were only four participants who finished the
task perfectly. Those, who did not accomplish the task, missed the
step that required them to `push the bobbin pin to the right' (the
$3^{\textrm{rd}}$ step of the sewing machine task in Table
\ref{tab:TaskExplanation}) as they did not clearly see the action
playing on the video guides.
These comments already hint to our future work on monitoring step state, however the results of overall performance are already encouraging and above 85\%.

The average completion times in the oscilloscope task are at
67.14($\pm$9.67) in the automatic video guide authoring condition and
67.00($\pm$12.17) seconds in expert video guide authoring condition.

In the screwdriver task, $73.29(\pm20.30)$s is the average completion time 
in the automatic edited condition and $71.14(\pm22.26)$s in the expert edited condition. The sewing machine task receives the highest completion time compared to the other two tasks at $134.86(\pm48.22)$ seconds in the automatic authoring condition, and $123.00(\pm34.93)$ seconds in the expert authoring condition. The completion time results also indicate the difficulty of each task.

In the oscilloscope task, the average numbers of video guides were played are $4.00(\pm1.00)$ and
$3.86(\pm0.38)$ videos in the automatic and the expert video guide
authoring condition, respectively.
In the screwdriver task, the average numbers of video guides are
similar to those of the oscilloscope task at $3.86(\pm0.69)$ videos in
the automatic authoring condition, and $4.00(\pm1.15)$ videos in the
expert authoring condition. The sewing machine task, however, has the
highest results of the numbers of video guides at $6.14(\pm1.95)$ and
$4.71(\pm0.76)$ videos in the automatic and expert authoring
condition, respectively.


One-way ANOVA was used to evaluate the completion time between
the automatic and the expert condition, since each user only performed
one condition in each task.

Figure~\ref{fig:CompletionTime} presents the pairwise boxplots of the
completion time in each task. In all tasks, there is no statistically
significant difference between the automatic and expert video guide
authoring conditions, i.e. ($t(12)=0.024,p=0.981$) in the oscilloscope
task, ($t(12)=0.188,p=0.857$) in the screwdriver task, and
($t(12)=0.527,p=0.617$) in the sewing machine task.

%
%


Results demonstrate that the video guides generated
using the model of attention are equivalent to the video guides manually
extracted by the experts. The user equally understands the automatically
generated video guides and is able to repeat the task.

The results obtained via this objective evaluation follow the impressions captured by the NASA-TLX scores which are omitted for brevity.


\subsubsection{The Participant's Opinions}

Participants' feedback was also collected and actual quotes are presented in
Table~\ref{tab:VideoAuthoringFeedback}. These are separated into
positive and negative feedback as well as additional opinions.

In the positive feedback, the participants overall
felt that the video guide was convenient and easy to follow and highlighted they did not require an instruction manual and could perform the tasks by just following the video guide. For the negative feedback, some participants found the video guide was too small and quick and one participant was confused by the expert's ego-motion in the video guide. In the screwdriver and sewing
machine tasks, the expert's hand covered parts of the object which blocked the view of details and caused them to skip one step in those tasks. Furthermore, the number of steps and details of the task affected the participants and prevented them to unable to finish every step of the task. As shown in the sewing machine task, only 4 out of 14 participants were able to perform the task perfectly.

Overall feedback shows that current implementation of GlaciAR is good for tasks with three or four steps. More complex tasks or tasks with small details may cause the user to skip some steps and not finish the task completely. Although showing the video guide from the expert's point of view made it easy for the novice user to understand the task, the expert's hand blocked the view in some crucial steps. For example, in the sewing machine task, the expert's thumb limits seeing how the bobbin aligns with the needle.

\begin{table}[h!] 
\centering
\begin{tabularx}{\columnwidth}{|l|X|}
\hline
Positive:
  & \textbullet {\it ``It is convenient and the user does not need
  the instruction.''}\\
  & \textbullet {\it ``I do not need to read the manual. Instead,
  I just watch and follow the video.''}\\
  & \textbullet {\it ``It is quite easy to follow the
  instructions.''}\\
  & \textbullet {\it ``Experts don't need to be hired to teach
  the task.''}\\
  & \textbullet {\it ``Hands-free application.''}\\
  & \textbullet {\it ``It is easy to follow the instruction, and
  the videos are slow enough to follow and understand.''}\\
  & \textbullet {\it ``It is easy to learn and follow the
  instruction.''}\\
\hline
Negative:
  & \textbullet {\it ``The screen is too small and I cannot see
  the video clearly, especially small buttons.''}\\
  & \textbullet {\it ``The video does not cover the whole area of
  the workspace, and is too fast.''}\\
  & \textbullet {\it ``Not much detail shown in some tasks.''}\\
  & \textbullet {\it ``Some details may be ignored or missing.''}\\
  & \textbullet {\it ``Some views make it quite difficult to see the
  details.''}\\
  & \textbullet {\it ``The expert's ego-motion in the video makes
  me watch the video several times to understand the task.''}\\
  & \textbullet {\it ``The experts' hands cover details in the task
  and make me skip those bits.''}\\
  & \textbullet {\it ``It is uncomfortable as I have to gaze at
  the screen.''}\\
  & \textbullet {\it ``The screen is a bit small.''}\\
\hline
Additional:
  & \textbullet {\it ``Lab demonstrator can be benefit from this
  device.''}\\
  & \textbullet {\it ``Instruction manuals for electrical and
  electronic devices.''}\\
  & \textbullet {\it ``Cooking and preparation.''}\\
  & \textbullet {\it ``Task reminder.''}\\
  & \textbullet {\it ``Tasks with not so many steps to follow.''}\\
  & \textbullet {\it ``Instructional videos similar to YouTube
  and how-to videos.''}\\
  & \textbullet {\it ``Training simple tasks to inexperienced
  users who don't speak the language written in the instructions.''}\\
  & \textbullet {\it ``DIY, especially simple tasks.''}\\
  & \textbullet {\it ``It does not need experts to instruct and 
  is easier to understand with the visual examples.''}\\
\hline
\end{tabularx}
\caption{Actual quotes from participants after using GlaciAR.}
\label{tab:VideoAuthoringFeedback}
\end{table}

\section{Discussion and Conclusion}

In this paper, the traditionally employed eye-gaze attention model is
replaced by a head-motion attention model. By combining this attention
model with a lightweight object detector, the system is able to
determine instances when users are paying attention, and then to use
this information to automatically edit video guides as well as
identify the moments in which users need guidance. The concept is
simple and capable of running on relatively low feature eye-wear
hardware without further user intervention.

The results indicate that GlaciAR, without any explicit user prompt, 
can guide users to perform a task that they may have never performed before.
The video guides are automatically extracted and
offered at the moments when the user walks up to an object of
interest. All three evaluated tasks were successfully accomplished in at
least $85.71\%$ of the tested cases.

Looking at, for example, the case of the oscilloscope, the object had
23 buttons/knobs and 7 sockets in an area of just $26\times13$ cm.
This relatively complex object had never been used before by any of
the volunteers, yet all of them were able to complete the task
successfully. A task with this level of choice in options could have been considered ideal
for other types of MR/AR formats such as those using 3D overlays.

It is thus somewhat surprising, yet encouraging that users were able to
achieve the task with the small images on the Glass screen. In addition,
the sensors and hardware used in GlaciAR are nowadays commonly available in mobile systems. GlaciAR uses
an IMU, a small 2D display, and low computational visual requirements.
Furthermore, as its operation requires no manual authoring and no synthesis
or labelling, the results of the video guide authoring evaluation also
suggest that the video guide produced by GlaciAR is as good as the one
edited by experts. The results show that, for every task tested, there is
no significant different between the two video guide conditions.

The feedback from both tests offer helpful suggestions for extending
the system towards a more practical one, such as by addressing the
lack of pause or rewind control over the video guide, the size of the
video due to the screen size, the lack of indicators to make the
starting or ending points of the video guides. Despite these
suggestions, there are also positive comments given by the
participants, such as that the intuition of the video guide makes it
easy to use.

In terms of directions for improvement, some of the above points would
be relatively easy to incorporate, but others would require an
additional strategy. The participants, for example, reported that the
view in the video guide was blocked by the expert's hand, and people
with glasses had problems watching the screen. Furthermore, the video
guides for more complex tasks need to have strategies to help the user
better follow the workflow, as demonstrated in the case of the sewing
machine task. Recent work has proposed mechanisms to model and keep
track of the workflow~\cite{Gabriel2015,Petersen2012,Petersen2013,cognitoPLOSOne}. One approach, for example, that would improve this issue, is a mechanism to evaluate the user's actions and workflow and present
any missing steps to the user to allow the user to finish the task
properly.

GlaciAR aims towards that elusive concept of MR/AR smart eye-wear
through which people can receive guidance to do any task anywhere and hence enabling cognitive augmentation.
This concept poses important challenges for conventional MR/AR systems,
especially due to the problem of authoring and scalability. GlaciAR shows encouraging high performance
 on three small-staged tasks and the discussion and results point to ways to improve the approach presented.


Further illustration on the way GlaciAR works is found on the accompanying video.
\section{Acknowledgments}
We are grateful to: 
%
\bibliographystyle{abbrv}
\bibliography{GlaciAR-arxiv}  
%
%
\end{document}